# Are gene-by-environment interactions leveraged in multi-modality neural networks for breast cancer prediction?

Monica Isgut[1,2], Andrew Hornback[2,3], Yunan Luo[3], Asma Khimani[1], Neha Jain[1], May D. Wang[2]

[1]Department of Bioinformatics, Georgia Institute of Technology, Atlanta, GA 30332, USA

[2]School of Biomedical Engineering, Georgia Institute of Technology and Emory University, Atlanta, GA 30322, USA

[3]School of Computational Science and Engineering, Georgia Institute of Technology and Emory University, Atlanta, GA 30322, USA

## Abstract

**Background:** Polygenic risk scores (PRSs) can significantly enhance breast cancer risk prediction when combined with routinely used clinical risk factor data. While many studies have explored the value-add of PRSs, little is known about the potential impact of gene-by-gene or gene-by-environment interactions towards enhancing the risk discrimination capabilities of multi-modal models combining PRSs with clinical data.

**Methods:** In this study, we integrated data on 318 individual genotype variants along with clinical data in a neural network to explore whether gene-by-gene (i.e., between individual variants) and/or gene-by-environment (between clinical risk factors and variants) interactions could be leveraged jointly during training to improve breast cancer risk prediction performance. We benchmarked our approach against a baseline model combining traditional univariate PRSs with clinical data in a logistic regression model and ran an interpretability analysis to identify feature interactions.

**Results:** While our model did not demonstrate improved performance over the baseline, we discovered 248 (<1%) statistically significant gene-by-gene and gene-by-environment interactions out of the ~53.6k possible feature pairs, the most contributory of which included rs6001930 (*MKL1*) and rs889312 (*MAP3K1*), with age and menopause being the most heavily interacting non-genetic risk factors. We also modeled the significant interactions as a network of highly connected features, suggesting that potential higher-order interactions are captured by the model.

**Interpretation:** Although gene-by-environment (or gene-by-gene) interactions did not enhance breast cancer risk prediction performance in neural networks, our study provides evidence that these interactions may nonetheless be leveraged by these models to inform their predictions. This study represents the first application of neural networks to screen for interactions impacting breast cancer risk using real-world data, highlighting their potential as tools for discovery.

**Funding:** No funding sources were reported for this study.

## Introduction

Polygenic risk scores (PRSs) have emerged as promising tools for enhancing breast cancer risk prediction in clinical settings. Established clinical risk models such as the Gail [1] or Tyrer-Cuzick scores [2], which calculate an incident breast cancer risk score based on non-genetic risk factors such as family history or age of menarche, are routinely used in practice to identify the most at-risk individuals who might benefit from preventive interventions such as breast magnetic resonance imaging (MRI) [3] or taking risk-lowering drugs like selective estrogen receptor (ER) modulators [4]–[6]. When considered alone or with basic demographic data, PRSs for breast cancer have been shown to achieve moderate risk discrimination capability, with published area under the curve (AUC) metrics ranging from 0.58 to 0.65 [3], [7]–[9], and with several scores capable of identifying individuals at over three-fold higher risk than others in the population [10], [11]. Many recent studies have combined established clinical breast cancer risk scores and/or other clinical risk factors with PRSs and demonstrated improved performance (i.e., "value-add") over clinical data alone, with improvements in AUC (i.e., $\Delta AUC_{PRS}$) ranging from around 0.02 to 0.08 for European ancestry populations [4], [9], [12]–[20].

A limitation of previous studies on the value-add of genomic data for breast cancer risk prediction is that they used univariate PRSs constructed using conventional approaches. PRSs for a given individual are conventionally calculated as a linear combination of weights (i.e., learned from genome-wide association studies; GWAS) and the individual's genotypes for a selection of variants, and are thus calculated as a pre-packaged score. When these univariate scores are integrated with clinical data in a multi-modal risk score, the individual weights associated with each genomic variant in the score are essentially fixed, thus precluding the ability of a risk model to simultaneously learn and adapt the importance weight attributions of the individual genomic variants together with the clinical features during training (**Fig. 1**). The potential opportunity cost of this is the inability to capture interactions between individual variants ("gene-by-gene" interactions) or between variants and clinical risk factors ("gene-by-environment" interactions), which may otherwise synergistically enhance disease prediction performance.

Gene-by-environment interactions have been extensively documented in the literature, but there have been limited studies exploring their potential utility for improving disease risk prediction [21]–[23]. Instead, much of the research on gene-by-environment interactions has been focused on identifying and characterizing these potential interactions [24]–[29]. Indeed, a standard approach towards exploring the usefulness of gene-by-environment interactions for disease prediction might involve specifying interaction terms ahead of time, such as by creating new composite features with individual genomic variables and one or more clinical variables (i.e., by multiplication), which would then be integrated into a linear model. Depending on the complexity of the feature interactions to be evaluated (i.e., pairwise, three-way, etc.) and the number of clinical and genomic variants in the model, this can become a computationally intractable approach and might best be done using a small feature set or with *a priori* knowledge of interactions of interest.

In this study, we address the gap in gene-by-environment and gene-by-gene interactions via neural networks. Neural networks serve as a promising alternative platform towards potentially leveraging interaction effects in risk prediction models to improve performance [30]–[35]. These models can capture complex input-output relationships and feature interactions by propagating data through a series of non-linear layers to produce the output. In this study, we designed a feature set comprising 318 individual genomic variants selected from two published polygenic risk scores [10], [36] (hereafter denoted as "PRS2") and then combined this with non-genetic risk factor data (denoted as "NG") to use as input into a neural network ("NN") for 10-year incident first-time breast cancer risk prediction. We evaluated model performance compared to its counterpart logistic regression ("LR") model using the same feature set (**Supplementary Fig. 1**), to determine whether the neural network resulted in improved performance. We also cross-checked these results with a baseline multi-modal model combining clinical features with two univariate polygenic risk score features (together denoted as "PRS1"), which were the same two published breast cancer PRSs that were used to select the PRS2 variants, as well as with single-modality versions of each feature set. The two PRS feature sets were designed using the same genomic variants, and the only difference between the two was the format in which the genomic variants were added to the model (i.e., pre-packaged as a fixed score, or as individual features; see **Figs. 1 and 2**). Finally, to gain a deeper understanding of the potential gene-by-gene or gene-by-environment interactions leveraged by the model, we ran an explorative interpretability analysis of the model to identify the most important features and their interactions, by adapting two highly cited neural network interpretation tools [37], [38].

Through this analysis, we aim to begin to explore whether and to what extent gene-by-environment (or gene-by-gene) interactions might be leveraged in neural networks to enhance disease prediction. The results of this study represent the first foundational exploration of gene-by-environment interactions for breast cancer in the context of a risk prediction neural network model, to our knowledge, and can serve as a starting point towards potential future use of neural networks to model these interactions.

**Results**

**Multi-modal integration of clinical data with individual genotypes in a neural network does not significantly improve breast cancer risk prediction performance over baseline models.**

We first sought to systematically compare the performance of the single-modality feature sets (PRS1, PRS2, and non-genetic) for 10-year first-time incident breast cancer risk prediction (**Table 1**, **Fig. 3a**). The non-genetic clinical risk factors (NG) dataset had the worst performance (AUC = 0.58 +/- 0.01 for both neural network and logistic regression models), and performance of the PRS1 and PRS2 single modality models was significantly higher (AUC = 0.61 +/- 0.02 for both feature sets and model complexity categories).

The AUC then increased further to 0.63 +/- 0.02 in each of the multi-modal neural network and logistic regression models containing both genomic and clinical features ("NG+PRS1" and "NG+PRS2"). While all multi-modal models achieved significantly better performance than the single-modality models, there was no significant difference in AUC between the multi-modal NG+PRS1 and NG+PRS2 models, regardless of whether they were trained using logistic regression or a neural network, and thus there was no difference in their value-add over models with just clinical data ($\Delta AUC_{PRS}$ from 0.05-0.06 +/- 0.02; **Fig. 3a-c** and **Table 1**). This suggests that the performance improvement attributed to the multi-modal models integrating genomic and clinical data was primarily driven by the main effects of the features in each data modality. Given that the models that included PRS2 features (i.e., 318 individual genotype variants) performed equivalently to their counterpart PRS1 models (i.e., two univariate polygenic risk scores), this suggests that there is no advantage in terms of AUC when variant weights are fixed versus learned simultaneously during model training.

Furthermore, as an auxiliary analysis, we then independently created a 3-covariate feature set comprising the two univariate polygenic risk scores (BC_77 and BC_313) in PRS1 with a third univariate polygenic risk score from published work comprising data from ~5k variants (BC_5k) [8]. When this model ("PRS1+BC_5k") was compared with and without also including clinical data, we found a higher AUC of ~0.65 and significantly higher value-add ($\Delta AUC_{PRS}$ ~= 0.07, Cohen's d = -0.09, p-value = 0.04) (**Fig. 3c**). While further studies are needed with additional polygenic risk scores other than BC_5k calculated using larger segments of the genome, this initial finding might suggest that the scope of genomic information used to generate a risk score may have a greater impact on model performance and value-add than the format in which genomic data are integrated into a risk model.

Interestingly, when we compared the raw score distributions (scaled between zero and one because of the sigmoid activation function) for cases and controls across the various models, we found that for the multi-modal neural network model that included the PRS2 dataset (NG+PRS2), the median raw score of the cases skewed slightly higher than that for all other models (**Fig. 3d**). The raw scores for cases for the multi-modal NG+PRS1 model also skewed higher compared to those for the counterpart single modality PRS1 and NG models, but to a lesser extent. Furthermore, the recall (i.e., ability to identify true cases) at various class decision thresholds over 0.5 was significantly higher for both multi-modal neural network models than for other models, especially for the NG+PRS2 model (**Fig. 3e**). For example, if the raw scores were to be used to classify at-risk individuals who might be targeted for primary prevention interventions, and if a class decision threshold (i.e., scaled score) of 0.8 were used as a cutoff, around 10-18% of true cases would be correctly identified with the NG+PRS2 neural network and the NG+PRS1 neural network would correctly identify 2-6% of true cases (**Supplementary Tables 1-2**). These findings were not observed in the logistic regression models (**Fig. 3e**), which had recall around or below 1% at the 0.8 decision threshold.

This improved recall only occurred at class decision thresholds over 0.5 and was at the expense of lower model precision at some of those class decision thresholds (i.e., 0.6, 0.7, and 0.8). However, at even higher-class decision thresholds (i.e., around 0.9), the multi-modal neural network NG+PRS2 model still achieved the best recall (~5% of true cases identified vs. <1% for all other models) but also had the highest precision (~8%) (**Supplementary Fig. 2**), suggesting improved risk discrimination at the upper extreme of the raw score distribution compared to the other models.

A similar trend occurs when comparing the single-modality genomic data models (PRS1 or PRS2 alone), where the PRS2 neural network model achieves significantly improved sensitivity at various decision thresholds above 0.5 compared its counterpart logistic regression model or the PRS1 models. For example, at class decision threshold 0.8 of the raw score, the PRS2 neural network model can identify ~5-14% of true cases, while the PRS1 neural network (BC_77 and BC_313 univariate scores combined) can identify around 0.1% (**Supplementary Tables 1-2**).

These findings imply that, despite all scores going through the same output (i.e., sigmoid) function during model training and being scaled between zero and one, the PRS2 and NG+PRS2 neural network models containing large-scale individual genomic variants as features rather than as pre-packaged univariate scores, have a greater overall range in the raw score outputs between the lowest and highest ends of the score distribution. This result is likely a reflection of the complexity of the function learned by these large-scale neural network models rather than a sign of improved sensitivity, given that there was no improvement in AUC or area under the precision-recall curve (AUPROC) compared to the other models. The results suggest that these models are learning a classification rule that strongly divides individuals into high and low scores, but that this rule is not significantly better at correctly identifying breast cancer cases than those of standard multi-modal models (i.e., it may be a local minimum in the loss function).

**Multi-modal models that integrated individual genotype features with clinical data enabled granular feature importance interpretation at the single-variant level.**

To begin to interpret these results, we then explored the feature importance rankings for each model, starting with the baseline multi-modal models with univariate PRSs (NG+PRS1). As shown in **Fig. 4a**, the most important features for the multi-modal NG+PRS1 logistic regression model (i.e., with clinical data and PRS1 features) were the two PRS1 polygenic risk scores (BC_313 followed by BC_77). For each feature, we ran a two-tailed one-sample t-test comparing the mean and standard error of its feature important weight (i.e., across the 10 trials) to zero, to identify features with importance weight magnitude significantly greater than zero. Along with the two PRS1 features, there were four significant non-genetic (NG) features (age, body mass index, and having a sibling or mother with a history of breast cancer) (**Fig. 4b**). When we compared the mean feature weights of this model with those of its neural network counterpart, both models were found to be highly correlated (Pearson $r^2 > 0.95$), and thus model complexity had no major impact on the feature importance trends for the NG+PRS1 feature set.

We next evaluated the feature importance of the multi-modal models containing the clinical features integrated with individual genotype variants as features (NG+PRS2). In the NG+PRS2 logistic regression model, there were 77 significant features, of which 70 (~91%) consisted of PRS2 features and the remaining 7 were from the clinical (NG) feature set. Of the significant PRS2 features, 11 (~16%) were variants selected from the BC_77 score, 53 (76%) were variants selected from the BC_313 score, and the remaining 6 (~8%) were variants that overlapped in both scores. Thus, most significant features were variants from the BC_313 score, which is concordant with the finding that BC_313 as a univariate score in the NG+PRS1 models was the feature with the highest mean importance weight. It seems that much of the performance value-add attributed to the PRS2 features is being driven by these BC_313 variants collectively. Interestingly, however, the significant BC_77 variants on average had significantly higher mean absolute value feature importance weights compared to any of the other feature types, and this was primarily driven by two outlier variants – rs6001930 (an intron variant in gene *MRTFA*, also denoted as *MKL1*) and rs889312 (an intergenic variant close to the *MAP3K1* gene) (**Fig. 4c, Fig. 4d**). This trend also occurred for the counterpart neural network models but to a lesser extent.

The mean feature importance weights of the NG+PRS2 neural network model are consistent with the counterpart NG+PRS2 logistic regression model results (Pearson $r^2 = 0.95$), suggesting a similar trend regardless of whether a linear or non-linear model was used for prediction. However, the neural network model only had 38 significant features (**Figs. 4e** and **4f**, **Supplementary Table 3, Supplementary Fig. 3**), suggesting potentially greater variability in model training per trial. This potentially reflects the challenges in effectively tuning neural network hyperparameters to ensure consistent performance. Nonetheless, the p-values of the significant features were highly correlated with magnitude of importance (Pearson $r^2 = 0.62$), suggesting overall consistency of those features across the trials (**Fig. 4e**). Out of the significant features in the NG+PRS2 neural network model, 24 (>60%) overlapped with the significant features in the counterpart multi-modal logistic regression model (**Supplementary Table 4**). Almost all significant features (~90%) were genomic variants. Out of the top 10 variants with the highest magnitude weights for each of the neural network and logistic regression NG+PRS2 models (total of 20), six of them (30%) overlapped, including rs6001930, rs88312, rs12406858 (intron variant in *TENT5C-DT*), rs10472076 (intergenic variant close to *RAB3C* and *PDE4D*), and rs4613718 (intergenic variant in chromosome 5) (**Table 2**). Other top-10 features in the NG+PRS2 neural network model included rs78540526 (intergenic variant in chromosome 11), rs625145 (intron variant in *SIK3*), rs1353747 (intron variant in *PDE4D*), and rs332529 (intergenic variant in chromosome 5). Interestingly, two of these variants on chromosome 5 were located close to the *PDE4D* gene and rs88312 near *MAP3K1*, was around 2 million base pairs away.

We then assessed the most important features for the PRS2 single modality neural network model. Most of these were intron variants in *FGFR2*, *ESR1* (two variants), *RAD51B*, *CASC16*, COL8A1, and *MRTFA* (the rs6001930 variant; the gene also denoted as *MKL1* or *MAP3K1*) (see **Supplementary Table 5**). When we then compared the feature importance trends for this model with that of the multi-modal NG+PRS2 neural network model, we found that that they were not correlated. The same was found between the multi-modal NG+PRS2 logistic regression model and its counterpart PRS2 logistic regression model. Just four variants were significant across all four single modality or multi-modal PRS2 models, including rs10941679 (intergenic variant close to *MRPS30*), rs17356907 (intergenic variant close to *USP44* and *PGAM1P5*), rs2588809 (intron variant in *RAD51B*), and rs78540526 (intergenic variant in chromosome 11 in an enhancer element), all of which were originally in both BC_313 and BC_77 scores. These results suggest that the integration of the clinical features with PRS2 variants seems to have an impact on the model training and loss trajectory, changing the scope of features used for disease prediction.

This was further supported by an additional analysis we did to compare the original published variant weights in the BC_77 and BC_313 scores to the mean feature importance weights learned by our trained single- and multi-modal models containing this feature set. The learned mean importance weight magnitudes for both PRS2 models, whether using a neural network or logistic regression, were moderately correlated with those of the original weights for BC_77 (Pearson $r^2$ = 0.48-0.49, p-value < 2 x $10^{-5}$), which supports the validity of the overall findings. However, there was no such correlation when the original weights for the same genomic variant features were compared against those learned by the multi-modal NG+PRS2 models. Given that this occurred for both the logistic regression and neural network NG+PRS2 models, it further strengthens the notion that integration of the non-genetic features had an overall impact on the loss trajectory of the model during training.

**Multi-modal neural networks combining individual genotypes with clinical data for breast cancer risk prediction provide insights into potential gene-by-gene and gene-by-environment interactions.**

To add further depth to this analysis, we explored the pairwise feature interaction effects in the multi-modal neural network models by adapting the highly-cited Neural Interaction Detection (NID) algorithm [38] after model training. As a baseline, we first explored feature interaction trends for the 66 feature pairs in the NG+PRS1 model. The mean interaction weight for each feature pair stratified by interaction type is visualized in **Fig. 5a**. To find interactions statistically likely to be "true", we ran a one-sample t-test for each feature pair, using the interaction weights obtained from NID for all ten trials for the pair, and then corrected for multiple comparisons. The distribution of the 31 significant interactions is shown in **Fig. 5b**. The pairwise interaction between the two univariate PRSs used in the model (BC_313 and BC_77) had the largest mean weight and was also the most significant (p-value = 5.28 x $10^{-8}$). This was followed by the NG-PRS ("gene-by-environment") interactions, and then the NG-NG interactions with the lowest mean weights (**Supplementary Table 6**). Most of the other largest-weight significant interaction pairs included either the univariate BC_313 or BC_77 scores as one of the features, including interactions with age, body mass index, and family history of breast cancer (**Figs. 5c and 5d**). Out of the top 15 highest-weight interaction pairs, 13 (87%) included either family history or at least one of the univariate PRSs as one of the features in the pair, suggesting a major contribution of both "gene-by-gene" and "gene-by-environment" interactions towards model performance. When interaction pairs were sorted based on significance (i.e., lowest p-values) in terms of their distance from zero, we observed the same general trend (**Figs. 5e and 5f**).

After this initial exploration of the baseline multi-modal model, we repeated this analysis for the multi-modal NG+PRS2 neural network model for the 53,628 unique possible pairwise interactions. Only 248 (<1%) feature pairs were significant, of which around 95% comprised PRS-PRS interactions (i.e., between two genotype variants), while just ~4% were NG-PRS interactions (i.e., between one genotype variant from either BC_313 or BC_77, and one non-genetic feature; denoted as NG-BC313 and NG-BC77 interactions) and less than 1% of significant pairs were NG-NG interactions, suggesting that the large-scale NG+PRS2 multi-modal model are most enriched for "gene-by-gene" significant interactions (**Figs. 6a and 6b**). Interestingly, we found a similar trend for the interactions that were not significant, where ~94% were gene-by-gene interactions, suggesting that this might be an artifact of the sheer number of PRS features (i.e., genotype variants), and many of them not interacting strongly with others. However, the significant variants were more enriched for interactions that included at least one BC_77 genotype variant, with ~67% of all significant interactions being of type BC77-BC77 between two genotype variants from the BC_77 score) or BC77-BC313 (between a BC_77 variant and BC_313 variant) versus just ~33% for the non-significant interactions. This might be attributed to BC_77 outlier variants, such as rs6001930 and rs889312, which interact strongly with other features. See **Supplementary Fig. 4** and **Supplementary Tables 6, and 8** for more details on the distribution of interaction weights and p-values for significant features.

Similarly to the NG+PRS1 multi-modal neural network, we explored two approaches towards interpreting the significant interactions in the NG+PRS2 neural network model. First, we ranked significant feature pairs from highest to lowest interaction weight to find those interactions likely to be contributing the most towards model output. Some of the highest-weight significant interactions in the entire model were enriched for feature pairs that included the BC_77 variants rs6001976, rs889312, rs135747 (an intron variant in the gene *PDE4D* coding for a phosphodiesterase), and/or rs10472076 (an intergenic variant close to *PDE4D* and *RAB3C*) (**Figs. 6c and 6d**, **Table 3**). Interestingly, the latter three variants are all located relatively close to each other on chromosome 5, within a span of around 2 million base pairs.

Other variants interacting strongly with either rs6001936 or rs889312 included rs7842619 (intron variant in *ANXA13* and *LOC105375739* in chromosome 8), rs12287832 (upstream variant near *OVOL1* in chromosome 11), rs55910451 (intron variant in *TASOR2* in chromosome 10), rs1172821 (intron variant in *BRSK1* in chromosome 19), and

rs7800548 (intron variant in *FAM185A*, *FBXL13* in chromosome 7), all originally from the BC_313 score, as well as rs78540526 (intergenic variant in chromosome 11 in enhancer element ; one of the 11 overlapping variants in both scores). More details on the top significant interactions sorted by mean weight are in **Table 3** and **Supplementary Table 9**.

When we then sorted the significant features based on lowest p-value (see **Supplementary Table 10** and **Supplementary Fig. 5, Supplementary Fig. 6**). Aside from the rs6001930 and rs889312 interaction, the most significant interactions were not necessarily the largest magnitude interactions, suggesting that these are less important for model predictions but nonetheless are may statistically be true interactions captured by the model. Interestingly, amongst the top five of most significant interactions, two of them included those between menopause and the variants, rs889312 and rs6001930, respectively, suggesting the potential relevance of this risk factor in "gene-by-environment" interactions.

The significant "gene-by-environment" (i.e., NG-PRS) interactions are described in **Table 4** and primarily included interactions with the risk factors menopause and age, as well as age of menarche, having had a hysterectomy, and frequent alcohol intake. The highest-weight and most significant NG-PRS interactions, respectively, were those between having had menopause and the rs6001930 variant (p-value = $1.58 \times 10^{-9}$, mean interaction weight = 0.61 +/- 0.09), and menopause with rs889312 (p-value = $1.35 \times 10^{-10}$, mean interaction weight = 0.55 +/- 0.06). Menopause and age participated in significant interactions with 3 and 11 different variants, respectively, and the interaction between menopause and age was the only significant NG-NG interaction in the NG+PRS2 model, further suggesting potential relevance of these risk factors.

Finally, we explored interpreting the significant pairwise interactions in a graph by representing all 248 features partaking in at least one significant interaction as nodes and the interactions themselves as edges. This is visualized in **Supplementary Figures 7 and 8**. Through this, our objective was to explore whether we could identify cliques of features that jointly strongly interact and to explore the overall connectivity of the features in relation to each other. We found a total of 101 maximal cliques (i.e., a clique such that, for each node, the clique is the largest complete subgraph containing the node). The largest two cliques consisted of nine features and included 6001930, rs1353747, rs889312, (the latter two located near each other on chromosome 5), and age, amongst other features (**Table 5, Supplementary Table 11**). Additionally, those three variants participated in the greatest number of maximal cliques (57, 47, and 37, respectively; **Supplementary Table 12**). Amongst the cliques that included "gene-by-environment" interactions with at least one non-genetic feature, an interesting one was that between age, menopause, rs6001930, and rs889312, providing further support to the potential relevance of these non-genetic features for model performance.

Overall, these results support the notion that these neural network models were mostly driven by "gene-by-gene" or epistatic interactions, but that some gene-by-environment interactions, primarily with age or menopause, are also relevant.

## Discussion

The goal of this study was to explore whether and to what extent gene-by-environment (or gene-by-gene) interactions might be leveraged in neural networks to enhance breast cancer risk prediction. We found that multi-modal neural networks combining 318 individual genotype variants with clinical risk factors did not achieve better performance over the baseline logistic regression multi-modal model combining two univariate PRSs with the same clinical data. Despite this, we found 248 significant interactions (<1%) out of the total of ~53.6k possible feature pairs in the model, providing evidence that while these interactions do not improve performance, they may still be leveraged by the neural network model to arrive at its predictions. Some of the most strongly interacting features contributing towards model performance included rs6001930 (*MKL1*) and rs889312 (*MAP3K1*), with the some of the most influential non-genetic features being age and menopause.

While several previous studies [4], [9], [12]–[20] have explored the value-add of polygenic risk scores when added to models with clinical data, these studies have used linear models and integrated PRSs in their traditional format, pre-packaged as univariate scores. The limitations of this approach are two-fold. First, real-world multi-modal data may not always be linearly separable, and interactions might exist between genomic data and environmental risk factors that may impact risk. Thus, using a linear model precludes the possibility for a risk score to potentially learn a more complex input-output function that might better identify the most at-risk individuals by leveraging non-linearities and interactions between inputs. Secondly, the standard format of a PRS is typically a univariate score calculated as a linear combination of genotypes and fixed weights. This precludes the ability of a multi-modal model including clinical data to jointly learn weights for the individual genotypes along with those for the clinical data, thus missing out on potential gene-by-gene (i.e., between genotype variants) or gene-by-environment (between genotype variants and clinical data) interactions that may otherwise enhance risk prediction performance. Our study is the first, to our knowledge, to integrate genotype data with clinical risk factors in a neural network for 10-year breast cancer risk prediction and to compare its performance to that of a linear model containing baseline PRSs constructed using the same genotypes but with fixed weights.

Both the baseline multi-modal logistic regression model with univariate PRSs (NG+PRS1; LR) and its neural network counterpart with individual genotypes (NG+PRS2; NN) achieved performance in line with previous studies, with AUCs of around 0.63 +/- 0.02, and high value-add over the models with clinical data alone ($\Delta AUC_{PRS}$ = 0.05-0.06 +/- 0.02). There are several possible reasons for this limited improvement, ranging from issues with the challenges of optimizing the training of a large-scale neural network or simply the lack of added improvement attributed to interaction effects. Our model interpretation analysis suggests the results were likely more a reflection of the former than the latter. Given the sheer number of parameters in the neural network and the likely small effect of each individual feature, the loss function trajectory was likely highly complex. While model performance was not worse than that of the baseline model, one of the smaller-importance significant features that should contribute towards increased risk (body mass index; BMI) was negatively associated with a high score, and other known non-genetic risk factors were not important for model performance (**Supplementary Fig. 3**). While we used both a random search and manual tuning to identify the optimal hyperparameters and model architecture for each trial, this model was also challenging to tune effectively. This strongly implies that the NG+PRS2 neural network was likely converging at a local minimum in the loss function during gradient descent. In other words, although it learned an effective classification rule comparable to that of the other models, it did not achieve its best-possible performance.

This notion is further supported by our finding that the range of raw scores for the NG+PRS2 neural network models was higher than that for some of the other models. This model might thus be learning a classification rule that strongly divides individuals into high and low scores but that is not significantly better at correctly identifying breast cancer cases than those of standard multi-modal models (i.e., it may be a local minimum in the loss function). Furthermore, the fact that both model sensitivity (i.e., recall) and precision were significantly higher than for other models at class decision thresholds of ~0.9 may implore the question of whether there might be a subset or sub-phenotype of cases that the model was identifying highly accurately, at the expense of misclassifying other cases and/or controls. Intriguingly, out of the 248 statistically significant pairwise interactions that we found in this model, some of the strongest interactions included variants for which there is some evidence in the literature of potential association with higher terminal duct lobular unit (TDLU) counts (rs11242675; *FOXQ1* and rs6001930; *MKL1*, respectively) and higher acini counts per TDLU (rs1353747; *PDE4D* and rs6472903; 8q21.11) [39]. Out of the significant interactions, rs6001930, rs135347, and rs11242675 are each part of 57, 47, and 16 maximal cliques, respectively, and are amongst the top-20 strongest interacting feature pairs in the model. Higher TDLU and acini counts are associated with certain breast cancer subtypes, such as the core basal phenotype (CBP; progesterone, estrogen, and human epidermal growth factor receptor negative) [40], which may support the notion that the multi-modal neural network models may be potentially arriving at a classification rule that effectively a subtype of cases.

Incidentally, unlike for other subtypes, evidence suggests that BMI is strongly linked with CBP risk in women under age 50 (p-value = 2 x $10^{-6}$) but is not associated with risk in women over age 50 [41]. A similar trend follows for progesterone receptor negative cancers, for which BMI is associated with a significantly reduced risk in women over age 50. This further concords with the interpretation of the NG+PRS2 neural network, for which the risk score is positively associated with post-menopause status but negatively linked to BMI, and for which age was one of the top interacting non-genetic features. While this potential explanation of model performance is purely qualitative and hypothesis-driven, it suggests potential value in future studies that aim to characterize the true positives and false negatives identified in these neural network models to explore whether they might be converging at a classification rule optimized for identifying certain breast cancer subtypes rather than all breast cancer cases.

Another interesting finding of this work is that, compared to the single-modality PRS2 models, the trend of feature importance for the genotype variants changed significantly when non-genetic risk factors were added (i.e., in the multi-modal NG+PRS2 models), regardless of whether this was using logistic regression or a neural network. For example, while the variant rs6001930 was the most significant feature in both multi-modal models, it was not significant in the single-modality PRS2 logistic regression model and was not the most significant feature in the PRS2 neural network model (**Supplementary Tables 4 and 5**). Furthermore, rs889312 (*MAP3K1*) was not significant at all in either of the single-modality PRS2 models but was the second most significant feature in both multi-modal models. The *MAP3K1* gene is a component of pathways such as the c-Jun-N-terminal kinase (JNK) pathway, which has been shown to be involved in inflammatory and apoptotic processes and to be activated by environmental factors [42]. One of the significant non-genetic features in both multi-modal models was post-menopausal status, which also had the most significant interaction with

rs889312 in the multi-modal neural network (**Supplementary Fig. 5**, **Supplementary Table 10**), suggesting potential value in future work aimed at exploring the potential role of estrogen levels and other post-menopausal biological changes in interacting with *MAP3K1* in its role in breast cancer susceptibility.

There are several limitations of this work. First, we used the neural interaction detection (NID) framework to identify interactions, and while this has been highly cited and has demonstrated strong performance at identifying true interactions (AUC from 0.79 to 0.98 depending on the dataset) [38], it is possible that there were false negatives and false positives in the set of interactions captured by the algorithm. While we used ten trials to arrive at a certain level of statistical confidence in the interactions, there is nonetheless the possibility that some interactions are not correct and other true interactions were missed by the model. Additionally, NID is limited in that it does not indicate the directionality or nature of any of the interactions, and thus the granularity of provided information is limited. Future work is warranted to establish a framework to validate whether the interactions captured in neural networks are true, i.e., real-world interactions, and to uncover the nature of these interactions with a greater level of granularity. This may involve running *in silico* mechanistic studies to explore protein-protein interactions or evaluate these findings in the context of known gene networks, or it may involve running simulation studies, or cross-validating the findings with other approaches designed for interpreting the interactions in a neural network, such as the framework proposed by Cui et al. which uses Shapley scores [30]. It may also involve simply adapting established statistical approaches (i.e., linear models, ANOVA) to validate the results. While we attempted the latter for just the top two variants, rs6001930 and rs889312, by multiplying them to create a composite feature, and did not find an interaction, it is possible that these interactions might exist only in conjunction with other features in higher-order cliques or as more complex functions than simple multiplication. Thus, approaches aimed at validating interactions found in neural networks should be flexible in the number of features and interaction terms evaluated.

Another limitation is in the neural network model optimization. Tuning the model hyperparameters was challenging, and it is possible, as discussed, that the model may be converging at a local minimum and could otherwise potentially achieve better performance. Advanced optimization algorithms, such as population-based search or Bayesian optimization could potentially be used to increase the chances that the model will arrive at the best possible input-output function given the data. Other modifications may be to adapt more bespoke modifications to the model architecture, such as using late fusion multi-modal integration approaches or skip connections, to facilitate training in a complex loss function landscape. Additionally, this analysis could benefit from inclusion of additional non-genetic breast cancer risk factors such as parity, breastfeeding information, or personal history of other cancers or diseases, which might potentially interact with genetic factors. Our study was also conducted on a European-only population, and future work is warranted that adapts these approaches towards other ancestry groups.

Our work represents the first-ever exploration of how gene-by-environment or gene-by-gene interactions might be leveraged in multi-modal neural networks for breast cancer risk prediction. While several simulation studies [31], [33], [34] have shown that neural networks hold promise for identifying such interactions, few recent studies have adapted these approaches towards real-world data [30], [35]. Existing gene-by-environment interaction studies for breast cancer have generally suffered from being underpowered, with discovered interactions often not reaching significance after correcting for multiple comparisons, and with findings not often being replicated in subsequent studies [24]. Our study was able to find significant interactions contributing towards model performance and may encourage the future use of neural networks to identify gene-by-gene or gene-by-environment interactions that affect breast cancer phenotypes or subtypes. In future work, neural networks may also be adapted to better elucidate mechanistic implications of any interactions. For example, genotype variants might be combined in a multi-modal model for breast cancer classification (i.e., versus risk prediction) along with more granular phenotype data than basic risk factor information, such as tumor imaging data or blood biomarker levels for cancer cases and controls, to arrive at a deeper mechanistic understanding from exploring feature interactions.

In summary, our study presents a first analysis to our knowledge of whether gene-by-environment or gene-by-gene interactions are leveraged in multi-modal neural networks with individual genotype variants for breast cancer risk prediction. We found potential higher-order and pairwise interactions, but model performance did not significantly improve. Thus, while these models may not be translationally more useful than existing baselines, this suggests potential utility of neural networks as a discovery tool for screening for interactions that impact breast cancer risk.

# Methods

## Overview of Approach

In this study, we designed a multi-modal feature set ("NG+PRS2") comprising clinical (i.e., non-genetic; "NG") breast cancer risk factors and individual genotypes for a selection of variants from two established polygenic risk scores ("PRS2") and evaluated the performance of this dataset compared to a baseline multi-modal dataset ("NG+PRS1") comprising the same clinical risk factors and variants, but with the variants pre-packaged into univariate polygenic risk scores using published fixed weights for each variant ("PRS1"). These feature sets were compared using logistic regression and neural network models. We also compared the performance of the single-modality feature sets (NG, PRS1, PRS2). Given that this study is replicated using both logistic regression and neural network models as two categories of model complexity, for consistency, we denote neural network-based 10-year incident breast cancer prediction models as "LR" and neural network-based 10-year incident disease prediction models as "NN".

In addition to benchmarking the various models, we also explored the interpretability of the four multi-modal models (NG+PRS1 and NG+PRS2, each for LR and NN). We first ran a feature interpretation analysis by comparing the learned input feature weights from the logistic regression models and by using the highly cited integrated gradients framework introduced by Sundarajan et al. [37] to compare the importance of each feature for the neural network models (**Supplementary Methods**). We repeated this for the single-modality models as a comparison. We then explored the feature interactions captured by each of the neural network models (NG+PRS1; NN and NG+PRS2; NN) to identify potentially meaningful gene-by-gene (between individual variants) or gene-by-environment (between variants and non-genetic risk factor features) interactions.

## Cohort Selection and Study Population

We utilized the UK Biobank dataset [43] comprising questionnaire results, physical measurements, diagnoses, and other clinical information for ~502,000 individuals. Data in the UK Biobank were collected during assessment visits in which a patient would attend a research site in person, and through online questionnaires or continuously updated national databases (i.e., death or cancer registry, or health records). All participants attended their first assessment visit, which occurred between 2006-2010, and all non-genetic features were data collected during each participant's first assessment visit. Participants with no medical history prior to their first assessment visit, with no available genotype data, or who opted to be removed from the UK Biobank were excluded from the analysis. To avoid potential ancestry-related confounders when analyzing the PRSs, our analysis centered on self-reported White British individuals only. We only included biological female individuals in the study.

To define the 10-year first-time breast cancer incident case and control cohorts, we used diagnostic codes available in the various diagnostic datasets available in the UK Biobank (i.e., primary care, hospital inpatient, cancer registry, etc.). We used each participant's year of recruitment into the UK Biobank as that participant's baseline time point, and for each disease, we excluded all participants with a pre-existing diagnosis of breast cancer prior to their recruitment. The 10-year window for each participant comprised the 10 years following their recruitment into the UK Biobank. Further details on the ICD-10 codes used to define cases for each disease are in **Supplementary Tables 13 and 14**, and the 10-year incident case and control counts are in **Supplementary Table 15**. The final sample included 5,174 incident cases comprising 2.68% of the ~193k participants.

## Preparation of Non-Genetic (NG) Risk Factor Feature Set

We used scientific literature to identify clinical risk factors associated with breast cancer and extracted proxy UK Biobank data fields containing representative data for each risk factor. We included risk factor features with large enough sample sizes and pre-processed the data by one-hot-encoding any categorical data fields as binary features and imputing any missing numerical values. For some of the features extracted from self-reported questionnaire data, some of the options included "prefer not to answer" or "unknown". We used these as control features to retain the full scope of data from each questionnaire. Some of the features, specifically sibling or maternal history of breast cancer can be proxies for genetic risk and thus is not strictly "non-genetic", but these features were included in the NG dataset, as this information is often included in clinical breast cancer risk scores. Further details on the clinical risk factor features and their pre-processing approach are in **Supplementary Table 16**.

**Preparation of PRS Feature Datasets**

For the baseline PRS feature set ("PRS1"), we identified breast cancer specific univariate PRSs that have previously been published in the literature. Each PRS was then calculated as a linear weighted sum of allele dosage by genome-wide association study (GWAS) weight for each variant, in concordance with conventional approaches. To calculate a polygenic risk score for each individual, we used the Plink 2.0 software[44] with the UK Biobank imputed genotypes dataset containing 96 million variants in chromosomes 1-22. The PRSs in the original published works contained 77 [36] and 313 [10] variants, respectively, of which 71 (~92%) and 258 (~82%) were available in our dataset (see **Supplementary Tables 17 and 18**). For the purposes of this study, we named these scores "BC_77" and "BC_313", respectively. The rationale for using more than one PRS was to increase the scope of information available in the PRSs, to enable a more robust analysis[45]. The BC_77 and BC_313 scores were moderately correlated, but not highly correlated, suggesting that each PRS may contain information independently of the other (Pearson $r^2 = 0.56$). The combination of BC_77 and BC_313 as a two-feature set is coined PRS1 for the purposes of Analysis 1. Another univariate PRS that was considered for this was a published breast cancer PRS using ~5k variants as input[8]. This and scores calculated with larger numbers of variants were not included as part of PRS1 due to the challenges of integrating the large-scale individual variants in a multivariate neural network, which would be required for the PRS2 comparison. However, we calculated this score and used it as a part of an auxiliary analysis comparing its performance to that of the other univariate PRSs (**Supplementary Table 18** and **Supplementary Figure 9**).

The novel per-variant PRS feature set ("PRS2") was derived using the unique variants pre-selected from the BC_77 and BC_313 scores. When all available variants from these scores (i.e., the 71 variants from BC_77 and the 258 variants from BC_313) were combined, this resulted in a dataset of 318 total variants, comprising 11 overlapping variants present in both datasets, 60 variants in the BC_77 dataset, and 247 variants in the BC_313 dataset. Each variant was encoded as 0, 1, or 2 based on the number of copies of the reference allele and used as an independent feature in the PRS2 dataset, resulting in a 318-feature dataset. Feature counts for each single-modality and multi-modal model are in **Supplementary Table 17**.

**Experimental Design for Model Training and Evaluation**

These PRS1 and PRS2 datasets were then compared in terms of their overall performance independently and in multi-modal models, as well as in terms of their value-add over non-genetic data alone, using both logistic regression and neural network-based algorithms (**Fig. 1, Supplementary Fig. 1**). The 5,174 total 10-year incident breast cancer cases and the respective female control samples in the dataset (n = 188,043) were then randomly split into train, validation, and test sets ten times (ratio of 70/20/10), and each random split comprised the dataset used for each of 10 trials for each model. As such, each train set contained around ~135k controls and ~3.7k cases. For each trial, random search-based hyperparameter tuning was done during training. For neural network models, hyperparameters included learning rate, input $L_1$ regularization, hidden layer $L_1$ regularization, input dropout, hidden layer dropout, number of hidden layers (from 1 to 3), and number of nodes per hidden layer. For logistic regression models, the hyperparameters were learning rate and $L_1$ regularization. Because of the challenges in training neural networks due to the larger hyperparameter space, we manually identified optimal hyperparameter ranges for these models prior to running the random search-based tuning. To account for class imbalance between cases and controls, we used a weighted loss function with the weight proportional to the case/control ratio for each of the trials. Models were trained for up to 50 epochs, and the model checkpoint at the best-performing epoch in the validation set (based on area under the curve; AUC) was used as the selected model for each trial. After training, the best-performing model for each trial was run on the respective test set for the trial (n~= 19.3k), and the mean and standard deviation of the performance metrics for the test set across the trials were computed. This was repeated for each feature set and model complexity category (i.e., NN and LR). Additional details are in the **Supplementary Methods**.

**Feature Interactions Analysis for Neural Network Models**

To explore whether any interactions between features in the neural network (M2) models (NG1+PRS; M2 or NG2+PRS; M2), we used a tool developed and validated by Tsang et al. (Neural Interaction Detection; NID) [38], which has been shown in simulation studies to be able to identify correct pairwise interaction effects with an area under the curve (AUC) ranging from 0.79 to 0.98 depending on the underlying structure of the data. We adapted their publicly available code to our multi-modal neural network models after initial training for each of the 10 trials. Using a pretrained model with learned parameters, the NID algorithm then returns a set of weights for each pair of features. NID is designed to provide two types of interaction analysis: a) pairwise interactions, and b) pairwise and higher-order interactions (denoted "any-order"). We explored focused on using the pairwise interactions analysis. Specifically, for each possible feature pair and for each node in the first hidden layer of a trained neural network model, NID calculates the interaction for that feature pair as the parameter with the lowest weight amongst the two features for that node. This is repeated for each node of the first hidden layer for the same feature pair, and the interaction weights for the feature pair are then summed across the nodes to arrive at a final pairwise interaction weight. This is then repeated for all other possible feature pairs.

We first analyzed potential interaction effects for the neural network trained on the baseline multi-modal feature set (NG+PRS1). Out of the 29 features in this model (27 NG and 2 PRS1 features), there were 406 unique feature pairs. Using NID, we calculated an interaction weight for each pair based on the respective joint importance of each feature in the pair in contributing towards the values of the first hidden layer neurons. Given that we had ten trials and a saved pretrained model for each trial, we repeated this for each trial.

To facilitate interpretability of the interactions, we represented each non-genetic risk factor as one feature related to each risk factor concept in the original dataset. For example, for "Menopause", we excluded any feature pairs containing the "No", "Prefer not to answer", and "Do not know" responses, and retained "Menopause – Yes", which we labeled as simply "Menopause". We merged "Sibling History of Breast Cancer" and "Mother History of Breast Cancer" into just one ("Family History") feature by using the average interaction weights across those features for each pair. We also merged "Alcohol Intake (3-4x per week)" and "Alcohol Intake (Daily or almost daily)" to one feature ("Alcohol Intake (Frequent)"). The final set of nine non-genetic features used in the interactions analysis is in **Supplementary Table 19**. This resulted in a total of 10 non-genetic risk factor features to compare, and a total of 66 unique interaction pairs for the NG+PRS1 model (10 NG feature concepts and 2 PRS1 features), including 20 NG-PRS interaction pairs, 45 NG-NG interaction pairs, and one PRS-PRS interaction pair (for BC_77 and BC_313).

We then scaled the interaction weights between zero and one for each trial and calculated the mean and standard error of the interaction weight for each of the 66 feature pairs across the trials. Using the mean and standard error interaction weight for each pair, we then established a metric of significance for each interaction pair weight by calculating its significance in terms of its distance from zero. The NID framework [38] is designed such that the interaction strength between any group of features is set to zero when an interaction does not exist at all, and the framework returned non-negative weights. Thus, we used a one-tailed one-sample t-test across the trials for each pairwise feature interaction in comparison against zero (i.e., to find pairs with significantly larger interaction weights than zero). We then used the Bonferroni correction to adjust for multiple comparisons across the 66 feature pairs (alpha = 0.01).

We repeated this approach for the multi-modal NG+PRS2 neural network model for the 59,340 unique possible pairwise interactions across the 345 total features (27 NG and 318 PRS2 features). Prior to scaling, we processed the non-genetic risk factor features in the same way as for the NG+PRS1 model, which resulted in an updated set of 53,628 interaction pairs across the remaining 328 features (10 NG feature concepts and 318 PRS2 features), including 3,180 NG-PRS interaction pairs, 20 NG-NG interaction pairs, and 50,403 PRS-PRS interactions (i.e., between individual variants). We then calculated the mean and standard error of the interaction weights, ran a one-tailed one-sample t-test to calculate the p-value results for the significance of each feature in its distance from zero, and then corrected for multiple comparisons based on the 53,628 possible interactions (alpha = 0.01).

For each neural network model, we then stratified feature pairs based on whether they were PRS-PRS interactions (just BC_313 and BC_77 univariate scores), NG-PRS interactions (a non-genetic feature and either BC_313 or BC_77), or NG-NG (two non-genetic features), to explore and visualize the results. For the NG+PRS2 model, we also stratified these pairwise interactions based on which published polygenic risk score each variant was originally selected from (i.e., from the univariate BC_77 or BC_313 publicly available list of variants, or from both).

To evaluate whether there might be a potential confounding effect of model architecture from using the NID approach, given that, for each feature pair, it returns a pairwise interaction weight value summed across all nodes in the first hidden layer, we compared this to a possible alternative approach in which we instead averaged the interaction weight across the nodes for each trial. This had no significant effect on the results (Pearson $r^2$ = 0.98 between the two approaches).

Finally, we also explored a graph-based interpretation approach to further visualize the significant interactions in the NG+PRS2 model. We created a graph using the significant pairwise feature connections and then evaluated the graph in terms of its cliques (i.e., groups of features that jointly interact) and subgraphs. This enabled more in-depth exploration of the interactions in the context of the network connectivity of the features.

Details on our additional analyses to explore these interactions are in the **Supplementary Methods**.

# Figures

**Figure 1: Illustration of differences between PRS1 and PRS2**

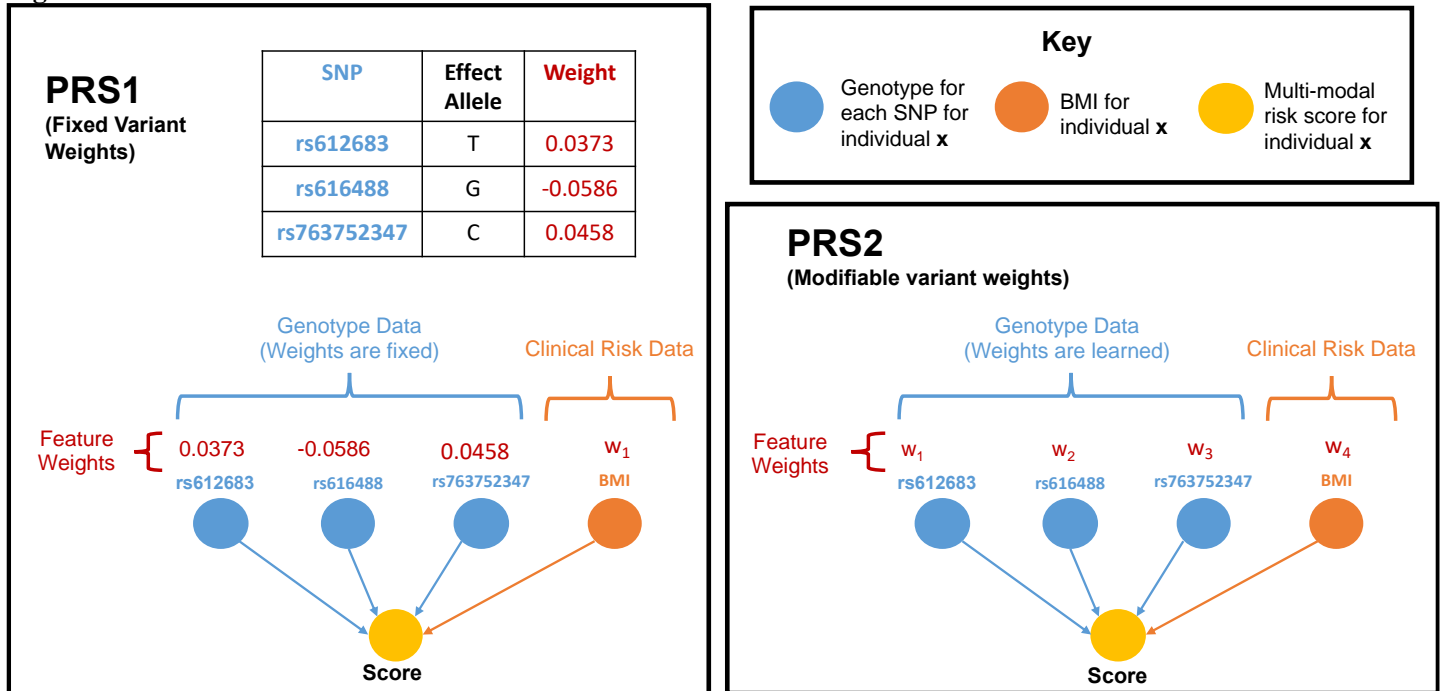

Abbreviations: SNP = single nucleotide polymorphism, BMI = body mass index, w = weight, NG = non-genetic risk factor. The multi-modal NG+PRS1 models use fixed weights for the genotype variants, pre-packaged in a linear combination, whereas the NG+PRS2 models learn the weights for the genotype variants simultaneously along with those for clinical risk data.

**Figure 2: PRS1 and PRS2 feature preparation**

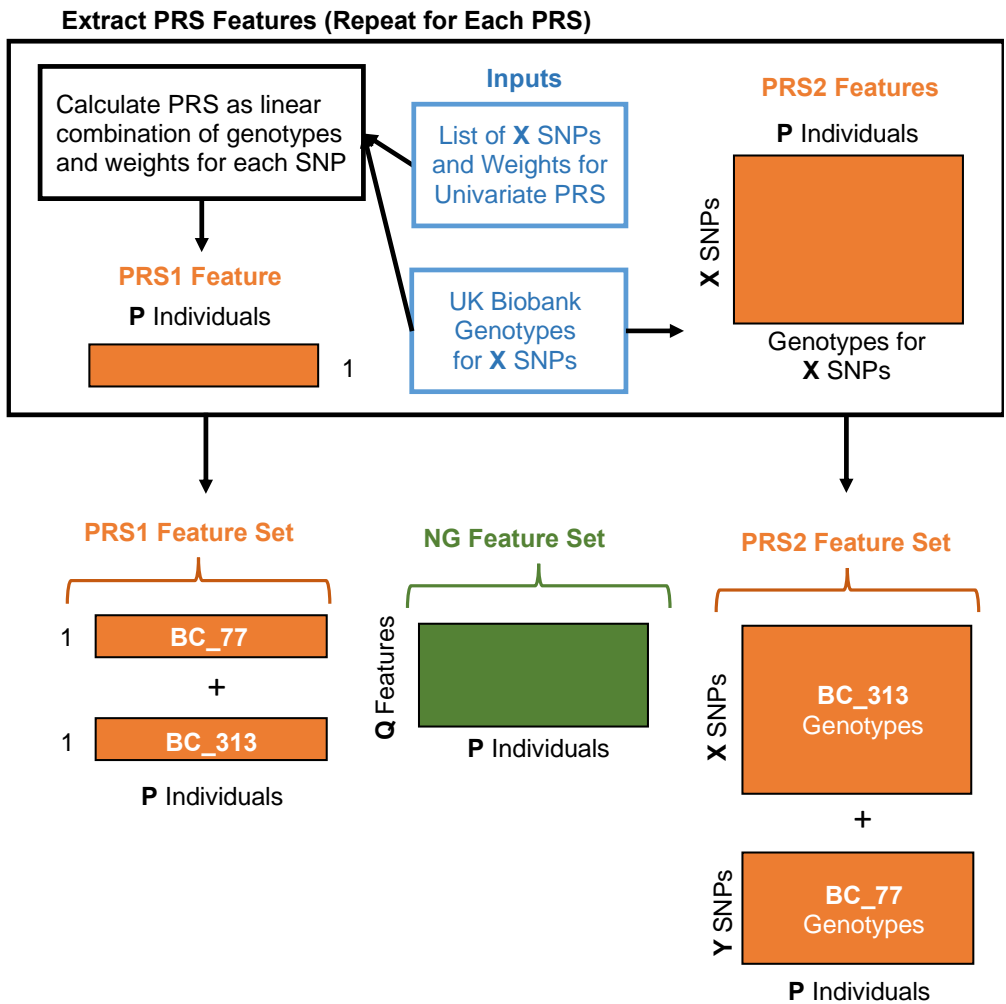

Abbreviations: PRS = polygenic risk score, SNP = single nucleotide polymorphism (also denoted as "variant"), NG = non-genetic, PRS1 = feature set with two univariate polygenic risk scores, PRS2 = feature set with multiple individual genomic features not pre-packaged as a univariate score. Variables: **X** = Number of SNPs in BC_313 polygenic risk score, **Y** = Number of SNPs in BC_77 polygenic risk score, **P** = Number of UK Biobank European ancestry individuals with available genotype data after basic sample filtering, **Q** = Number of selected clinical (i.e., non-genetic, NG) risk factors for breast cancer. Two published breast cancer polygenic risk scores (BC_77 and BC_313) were used for this analysis. From an external list of variants and published weights for each PRS, we used the UK Biobank genotypes dataset to calculate a score for each participant of European ancestry with available genotype data after basic sample filtering (PRS1). We also extracted the available genotype features for the variants in each PRS, combined them together, and removed any overlapping (i.e., duplicated) variants (PRS2). Samples were then split into train, validation, and test (70/20/10 ratio).

**Figure 3: Key performance results for neural network and logistic regression models**

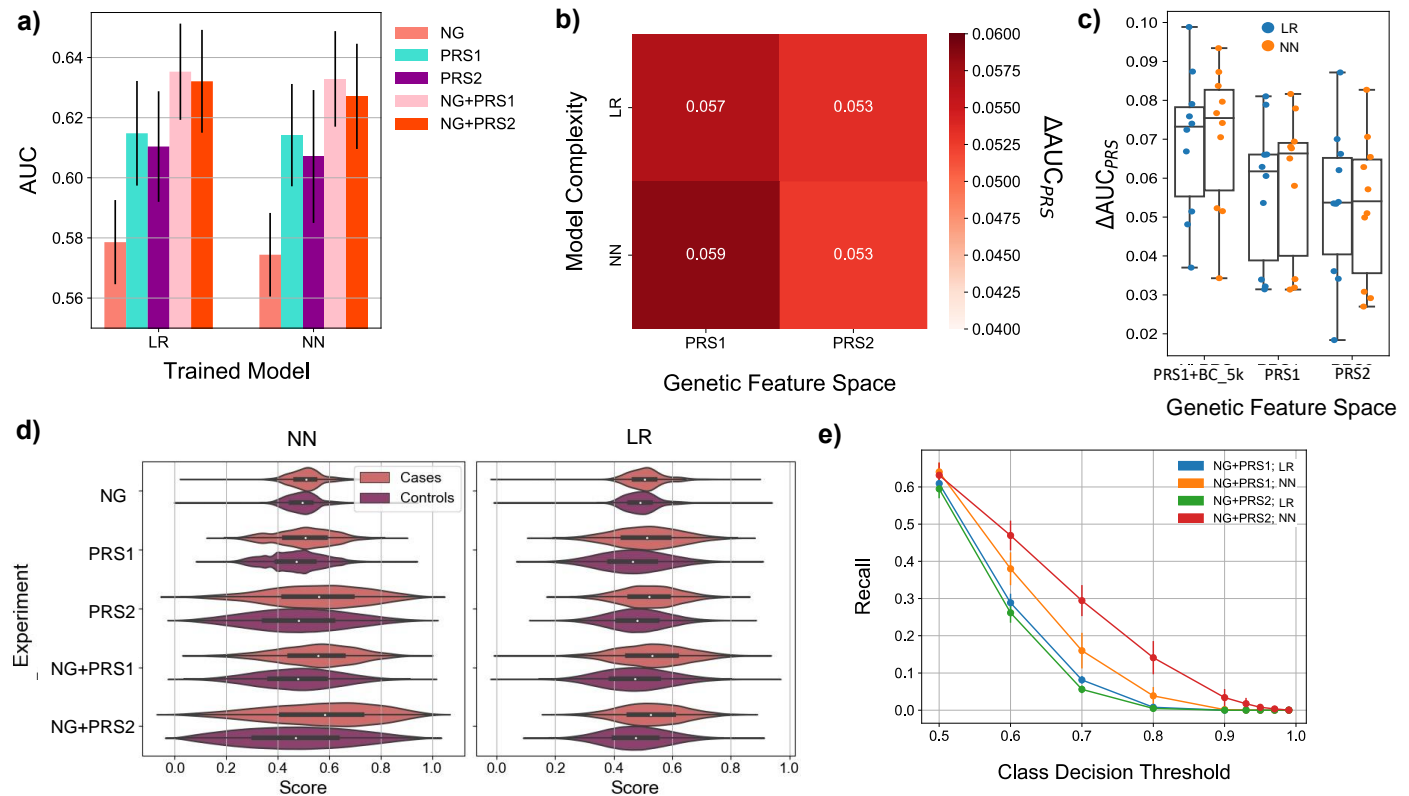

Abbreviations: AUC = area under the curve, PRS = polygenic risk score, NG = non-genetic (i.e., clinical) features, PRS1 = polygenic risk score category 1 (univariate polygenic risk score features), PRS2 = polygenic risk score category 2 (large-scale individual genomic features), M1 = model complexity category 1 (i.e., logistic regression), M2 = model complexity category 2 (i.e., neural network). The value-add of each polygenic risk score is calculated as $\Delta AUC_{PRS} = AUC_{NG+PRS} - AUC_{NG}$. **a)** Performance measured in AUC for each single modality or multi-modal model across 10 trials on the test set. PRS1 and PRS2 achieve similar performance. Multi-modal models achieve the highest performance, but there is no significant difference between combined NG+PRS1 and NG+PRS2 models. **b)** The mean value-add is shown as $\Delta AUC_{PRS}$ for each of PRS1 and PRS2 over non-genetic (i.e., clinical) data alone, benchmarked using logistic regression and neural network models across 10 trials on the test set. Both PRS1 and PRS2 feature sets improve the AUC by 0.05 over clinical data models alone but do not differ significantly in their value-add. **c)** The $\Delta AUC_{PRS}$ is shown as a boxplot to visualize the distribution across the categories of model complexity over 10 trials. PRS1 (including BC_77 and BC_313) and PRS2 were compared. The PRS1 scores were then combined with an established univariate breast cancer score developed using ~5k variants (BC_5k) to create a three-feature dataset, which was then compared and was shown to significantly improve prediction value-add of clinical data alone. **d)** Violin plots show the distribution of each score for 10-year incident breast cancer prediction for cases and controls for logistic regression (M1) and neural network (M2) models across all 10 trials. For the multi-modal NG+PRS2 model, the median score of cases (not controls) skews higher, but only when neural networks (M2) are used. The same is shown with the NG+PRS1 and PRS1 compared to the single modality models, but not to the same extent. **e)** The recall (i.e., ability to identify true cases) is shown for various class decision boundary thresholds for each multi-modal model. It is significantly higher for NG+PRS2 neural network (M2) models for some class decision thresholds (0.6, 0.7, 0.8, and 0.9).

**Figure 4: Feature importance for NG+PRS1 and NG+PRS2 multi-modal neural network and logistic regression models**

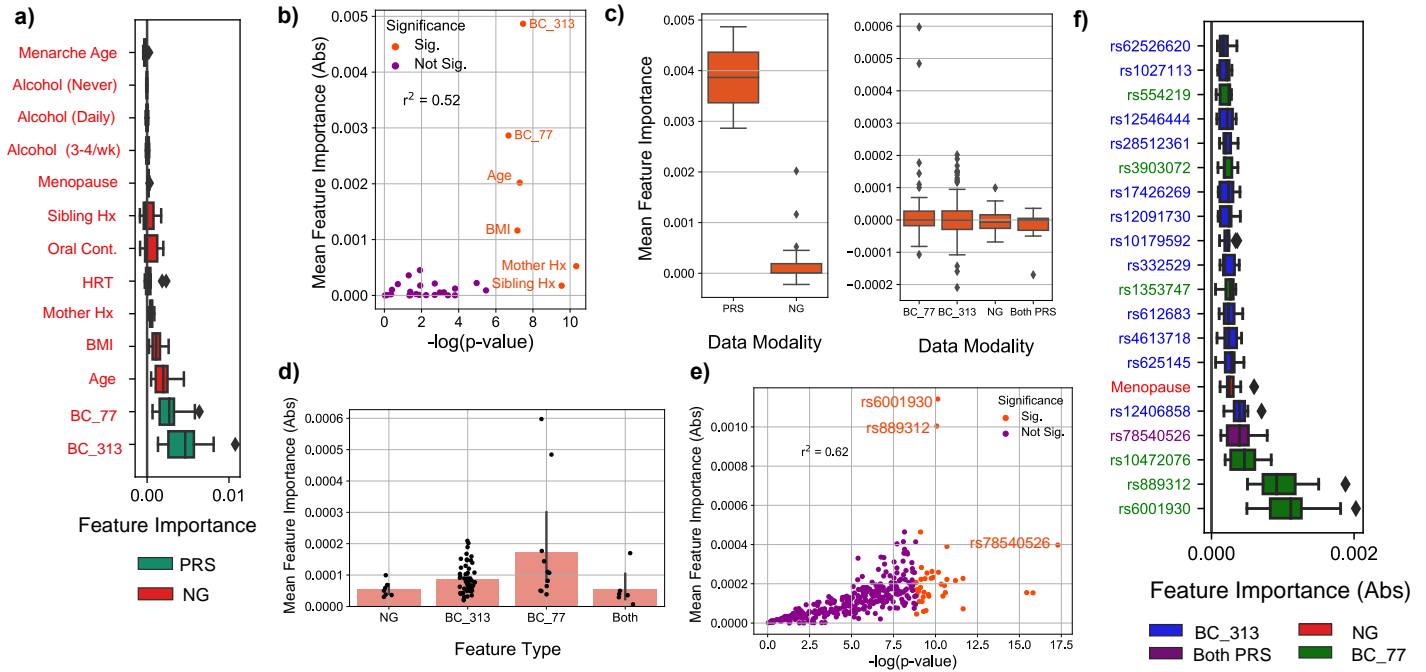

Abbreviations: HRT = hormone replacement therapy, Hx = history, BMI = body mass index, Cont. = contraceptive, wk= week, Sig. = significant, Abs = absolute value, NG = non-genetic (i.e., clinical) data, PRS = polygenic risk score. **a)** Boxplots of the distribution of importance (i.e., logistic regression input weight) for each feature in the multi-modal NG+PRS1 logistic regression model. The most important features contributing towards increased breast cancer risk were the two PRSs (BC_77 and BC_313), followed by non-genetic features including age, BMI, and having a mother with a history of breast cancer. We included several control features in the model (not included in this figure), and as expected, the importance of these features was not significantly different from zero. Age of menarche, as expected from the literature, was negatively associated with risk on average. For visualization purposes, we only included one feature for each risk factor concept. **b)** A scatterplot for the multi-modal NG+PRS1 logistic regression model showing the feature importance weight and –log(p-value) from a two-tailed one-sample t-test comparing the mean and standard deviation of the importance weight for each feature across 10 trials to zero as a measure of whether a given feature is significantly different from zero. Features in orange have importance weights significantly greater than or less than zero after adjusting for multiple comparisons using the Bonferroni correction at alpha = 0.05. While BC_313 and BC_77 have the highest mean absolute value feature importance weights amongst all features, the most significant risk factors with the lowest p-values were having a family history of breast cancer (mother and sibling), suggesting a lower effect but greater consistency amongst trials. Age and BMI were also significant features. We confirmed that the absolute value feature importance weight and –log(p-values) were strongly correlated ($r^2 = 0.52$) for the significant features as validation of the analysis. **c)** (Left) Mean feature importance weight for each data modality in the multi-modal NG+PRS1 logistic regression model, (Right) Mean feature importance weight for each data modality in the multi-modal NG+PRS2 logistic regression model. PRS2 features are positive or negative depending on the risk allele. "Both PRS" refers to the 11 variants that overlap between BC_77 and BC_313 polygenic risk scores. **d)** A bar plot of absolute value mean feature importance shown for each data modality in the NG+PRS2 logistic regression model. BC_77 significant features have higher absolute value weights than all other category features. **e)** A scatterplot is shown with absolute value feature importance weight and -log(p-value) for NG+PRS2 neural network features. The p-values were calculated using a one-sample t-test for each feature across 10 trials comparing it against zero. The Pearson correlation of 0.62 between -log(p-value) and feature importance weight absolute value for significant features supports the reliability of the findings across the 10 trials. **f)** Boxplots of the distribution of absolute value importance weight (calculated using integrated gradients) for the top 20 significant features with the largest magnitude weights in the NG+PRS2 neural network model. The variants, rs6001930 and rs889312, originally selected from the BC_77 score, have the largest feature importance contributing towards risk. Other top features are rs10472076, rs78540526, rs12406858, and menopause.

**Figure 5: Pairwise feature interactions for multi-modal NG+PRS1 neural network model**

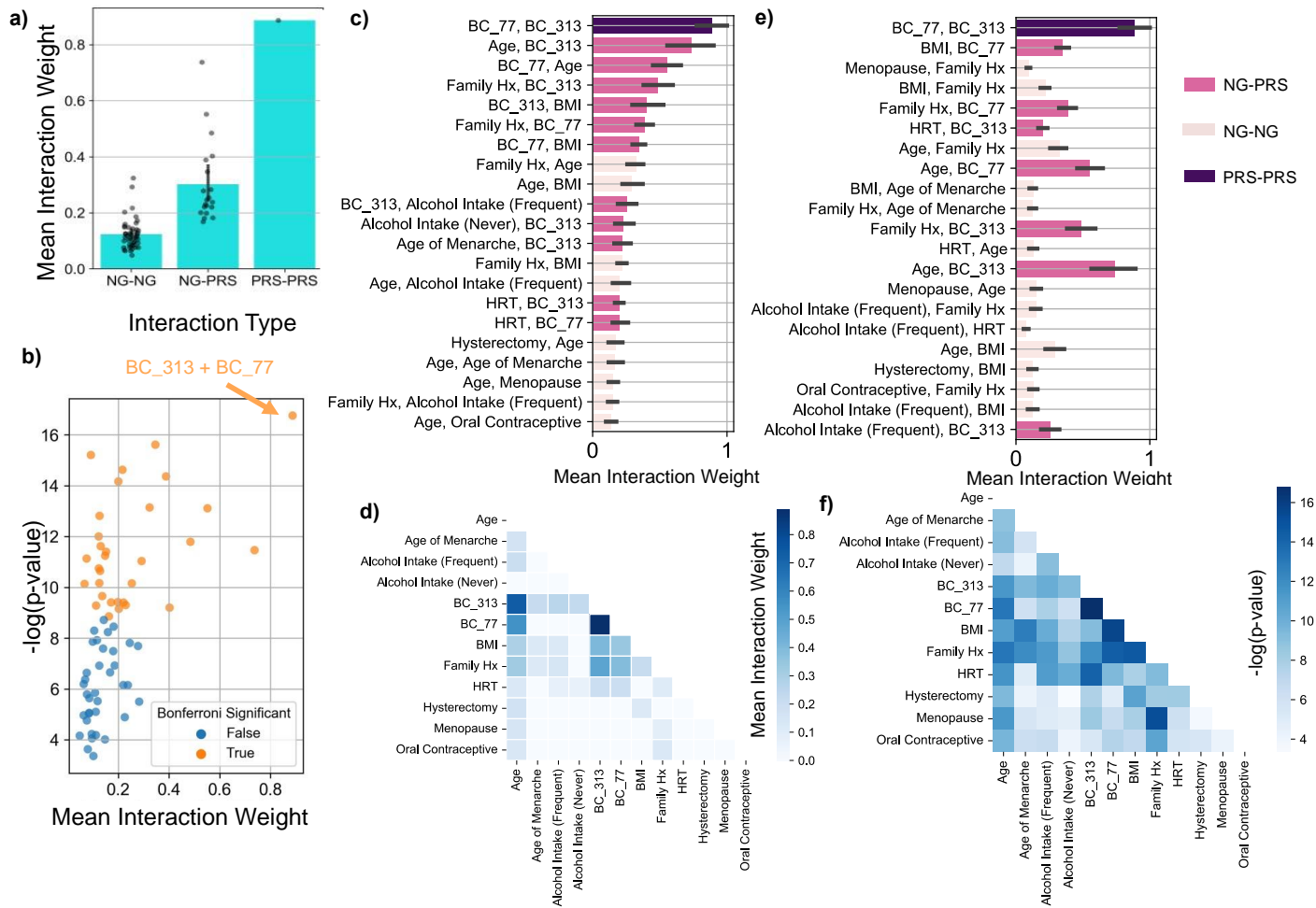

Abbreviations: NG = non-genetic (i.e., clinical) features, PRS1 = univariate polygenic risk score features (BC_77 and BC_313), PRS2 = individual genotype features, BMI = body mass index, Hx = history, HRT = hormone replacement therapy, NG-NG = interaction between two non-genetic features, NG-PRS = interaction between a non-genetic feature and a univariate polygenic risk score (BC_77 or BC_313), PRS-PRS = interaction between the two univariate polygenic risk scores, BC_77 and BC_313. **a)** A bar plot of the mean interaction weights for each type of interaction. The interaction between BC_77 and BC_313 is the one with the highest weight. The NG-NG interactions have the lowest interaction weights on average. **b)** A scatterplot of the mean interaction weight for each feature pair and the negative log of the p-value of significance for the feature pairs. Significance was calculated for each feature pair by using a one-tailed one-sample t-test to compare its per-trial weights to zero. The 31 significant interactions (i.e., those significantly greater than zero) are colored in orange. The interaction between the two PRSs is an outlier, with both the largest mean interaction weight and the lowest p-value. **c)** A bar plot with the top 25 significant interactions, sorted from high to low mean interaction weight across the trials. The points represent the interaction weight for each trial for each feature pair. Most of the most highly weighted interactions include at least one of BC_77 or BC_313, and family history, which is a proxy for genetic risk, is also amongst the top interacting features. Other features that interact strongly include age and body mass index. **d)** A heatmap with the mean pairwise interaction weights across features comprising the top 20 significant interactions with the largest mean interaction weights. Aside from the BC_313 and BC_77 interaction, other top interacting feature pairs include BC_313 and BC_77, respectively, each with both family history and age. **e)** A bar plot with the top 25 significant interactions, sorted from low to high p-value across the trials, with interaction weights for each trial shown as points. Other than BC_313 and BC_77, the most significant interactions are not necessarily the largest-weighted ones but may represent true interactions captured by the neural network model. **f)** A heatmap with the top 20 significant interactions sorted by lowest p-value. Some of these top interactions include those between family history and body mass index and menopause, respectively, and between body mass index and the BC_77 score.

**Figure 6: Pairwise feature interactions for multi-modal NG+PRS2 neural network model**

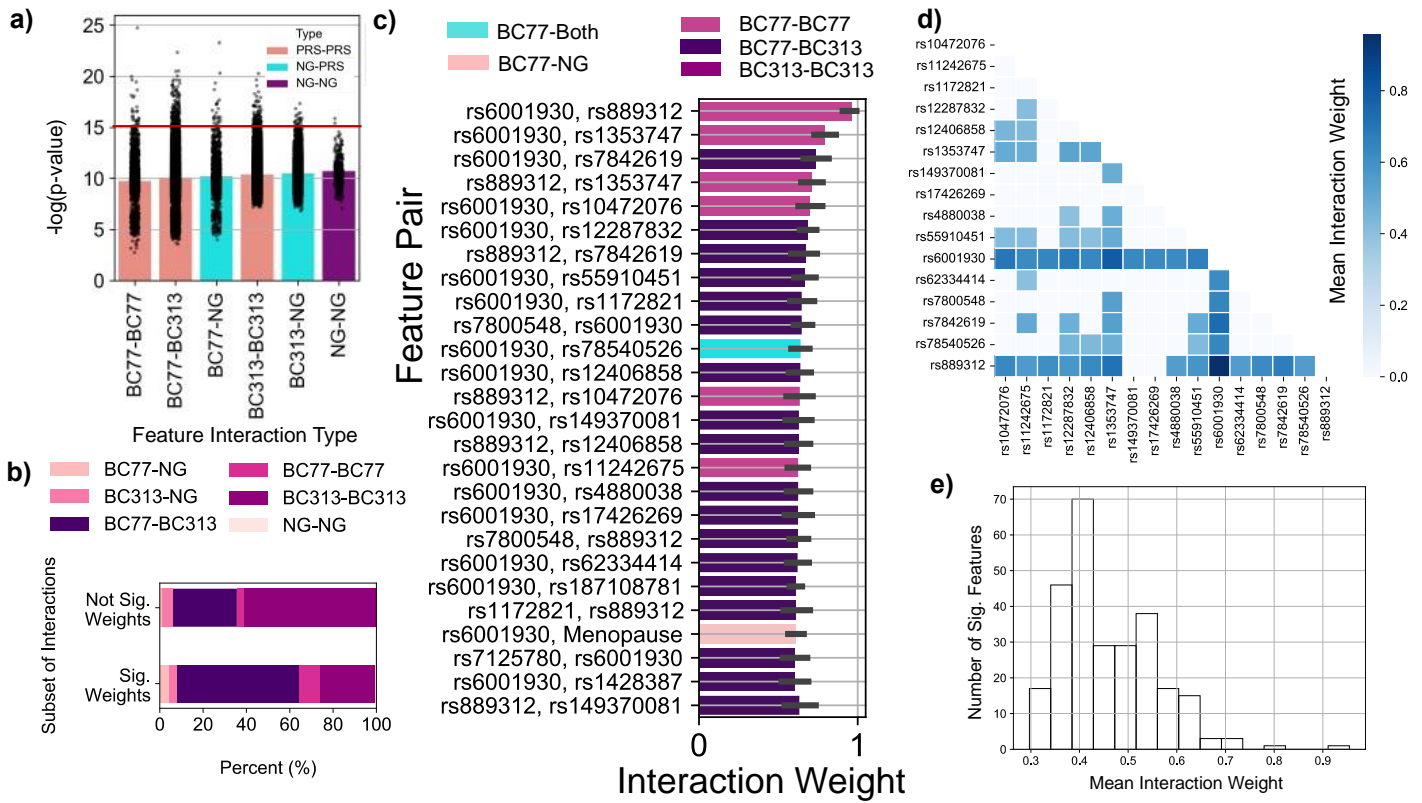

Abbreviations: NG = non-genetic, Sig. = significant, PRS-PRS = interaction between two genotype variant features, NG-PRS = interaction between a genotype feature and a non-genetic risk factor, NG-NG = interaction between two non-genetic risk factors. BC77-NG = interaction between one feature selected from the BC_77 polygenic risk score and one non-genetic feature, BC313-NG = interaction between one feature selected from the BC_313 polygenic risk score and one non-genetic feature, BC313-BC77 = interaction between one feature selected from the BC_313 score and one selected from the BC_77 score. Any interactions with "Both", such as BC77-Both, consist of interactions with at least one feature derived from one of the 11 overlapping variants in BC_77 and BC_313. **a)** Bar plot of significance p-values for all interactions stratified by type. The p-values were calculated using a one-sample one-tailed t-test for each interaction weight across the 10 trials in comparison against zero. Any features that were significantly greater than zero after correction for multiple comparisons (n = 53,628 interaction pairs, alpha = 0.01). The approximate negative log p-value threshold for significant interactions is shown. The most significant interaction is the BC77-BC77 interaction between rs6001930 and rs889312. While most significant interactions (>90%) are "gene-by-gene" (PRS-PRS), many are "gene-by-environment" (NG-PRS). **b)** A stacked bar plot showing the distribution of significant and non-significant interaction weights, respectively, by interaction type. The set of significant and non-significant interaction pairs comprise roughly the same proportion of NG-PRS (including BC77-NG and BC313-NG) and PRS-PRS interactions, however the significant interactions are more heavily dominated by interactions between one BC_77 variant and one BC_313 variant (~56% vs. ~29%) and are more enriched for BC77-BC77 interactions (~10% vs. ~4%). Amongst the NG-PRS interactions, the significant pairs than non-significant pairs are more likely to be of type BC77-NG (4.6% vs. 1.2%). This is likely a result of the high level of interaction between rs6001930 and rs889312 and other features. **c)** The top 25 pairwise significant interactions, sorted by mean interaction weight. **d)** Heatmap of mean interaction weight for the features within the set of top 20 pairwise interactions. **e)** The distribution of mean interaction weight for significant interactions. The weights range from ~0.3 to almost 1. Note: For subfigures a and b, to simplify the figure (i.e., to avoid using extra PRS-PRS interaction type categories), we excluded the ~3.5k interactions in which one or both features comprised one of the 11 variants derived from both BC_77 and BC_313 scores. For subfigure b, the percentages are based on this abridged set of ~51k feature pairs.

# Tables

**Table 1: Model performance for 10-year incident breast cancer prediction**

| Model Complexity | NG Features | PRS Features | | Multi-Modal Features | | PRS Value-Add | |
|:---:|:---:|:---:|:---:|:---:|:---:|:---:|:---:|
| | $AUC_{NG}$ (SE) | $AUC_{PRS1}$ (SE) | $AUC_{PRS2}$ (SE) | $AUC_{NG+PRS1}$ (SE) | $AUC_{NG+PRS2}$ (SE) | $\Delta AUC_{PRS1}$ (SE) | $\Delta AUC_{PRS2}$ (SE) |
| LR | 0.5786 (0.0140) | 0.6148 (0.0174) | 0.6104 (0.0184) | 0.6353 (0.0160) | 0.6321 (0.0171) | 0.0566 (0.0175) | 0.0535 (0.0188) |
| NN | 0.5744 (0.0139) | 0.6142 (0.0170) | 0.6071 (0.0221) | 0.6329 (0.0159) | 0.6271 (0.0175) | 0.0585 (0.0181) | 0.0527 (0.0179) |

Abbreviations: AUC = area under the curve, SE = standard error, NG = non-genetic features, PRS1 = polygenic risk scores version 1 (combined univariate scores), PRS2 = polygenic risk scores version 2 (individual variants), LR = logistic regression, NN = neural network. Performance results are shown for the test set across each of 10 trials. Results for all models, including those with only genetic features, are shown using case/control labels specific to 10-year first-time incident diagnosis prediction from the time of each participant's recruitment into the UK Biobank. The $\Delta AUC_{PRS}$ values are calculated through subtracting the $AUC_{PRS}$ result from the $AUC_{NG+PRS}$ for each of ten trials for each respective model complexity subcategory (LR vs. NN) and genetic feature space category (PRS1 vs. PRS2), and then calculating the mean and standard error across the trials.

**Table 2: Top 10 most important SNPs in each NG+PRS2 multi-modal model for 10-year incident breast cancer risk prediction (based on absolute value feature importance weight)**

| Model | SNP | Position | Original PRS | p-value | Type of Variant | Proximal Gene(s) |
|---|---|---|---|---|---|---|
| NN and LR | **rs6001930** | 22:40876234 | BC_77 | $3.94 \times 10^{-5}$ (NN) $2.42 \times 10^{-8}$ (LR) | Intron | *MRTFA* |
| | **rs889312** | 5:56031884 | BC_77 | $4.20 \times 10^{-5}$ (NN) $2.55 \times 10^{-7}$ (LR) | Intergenic | *LOC105378979, C5orf67, MAP3K1* |
| | **rs12406858** | 1:118141492 | BC_313 | $2.20 \times 10^{-5}$ (NN) $1.01 \times 10^{-6}$ (LR) | Intron | *TENT5C-DT* |
| | **rs10472076** | 5:58184061 | BC_77 | $1.11 \times 10^{-4}$ (NN) $1.59 \times 10^{-6}$ (LR) | Intergenic | *RAB3C, PDE4D* |
| | **rs4613718** | 5:44649944 | BC_313 | $4.36 \times 10^{-6}$ (LR) $3.81 \times 10^{-5}$ (NN) | Intergenic | N/A |
| | **rs78540526** | 11:69331418 | Both PRS | $2.08 \times 10^{-13}$ (LR) $3.10 \times 10^{-8}$ (NN) | Intergenic | N/A |
| NN | **rs625145** | 11:116727936 | BC_313 | $5.57 \times 10^{-5}$ | Intron | *SIK3* |
| | **rs612683** | 1:100880328 | BC_313 | $7.51 \times 10^{-5}$ | Intron and NC | *CDC14A, LOC105379827* |
| | **rs1353747** | 5:58337481 | BC_77 | $5.14 \times 10^{-5}$ | Intron | *PDE4D* |
| | **rs332529** | 5:90789470 | BC_313 | $5.56 \times 10^{-5}$ | Intergenic | N/A |
| LR | **rs7842619** | 8:124739913 | BC_313 | $1.54 \times 10^{-6}$ | Intron | *ANXA13, LOC105375739* |
| | **rs149370081** | 17:40744470 | BC_313 | $3.96 \times 10^{-6}$ | Intron | *RETREG3* |
| | **rs7800548** | 7:102481842 | BC_313 | $7.56 \times 10^{-7}$ | Intron | *FAM185A, FBXL13* |
| | **rs1172821** | 19:55816678 | BC_313 | $7.51 \times 10^{-5}$ | Intron | *BRSK1* |

Abbreviations: NN = neural network, LR = logistic regression, NC = non-coding sequence. The SNPs are in order of highest to lowest weight, all of them are significantly different from zero. Variants in the top 10 of both NN and LR models are included as "NN and LR" category. Note: The alternative name for *MRTFA* is *MKL1*.

The two SNPs rs6001930 and rs889312 show up as the most important for both multi-modality models, logistic regression and neural network. The variants heavily overlap and are correlated. However, these variants are also the features with the most amount of interaction in the neural networks. At the same time, all this interaction is not doing much in terms of improvement. Might be playing a role in the improved model recall. But the value-add seems to be mostly a result of the independent effect of the PRS2 features.

**Table 3: Top 25 significant feature interactions in NG+PRS2 neural network model with the largest mean weights**

| Feature 1 | Feature 2 | Mean Interaction Weight | Std | p-value | Feature 2 Description | Top-10 Most Important SNP |
|---|---|---|---|---|---|---|
| **rs6001930** | **rs889312** | 0.96 | 0.08 | 1.37E-11 | N/A | N/A |
| **rs6001930** | rs1353747 | 0.79 | 0.12 | 2.67E-09 | Intron variant in *PDE4D* in chromosome 5 | NN |
| **rs6001930** | rs7842619 | 0.73 | 0.14 | 2.52E-08 | Intron variant in *ANXA13*, *LOC105375739* in chromosome 8 | LR |
| **rs889312** | rs1353747 | 0.71 | 0.12 | 5.98E-09 | Intron variant in *PDE4D* in chromosome 5 | NN |
| **rs6001930** | rs10472076 | 0.70 | 0.13 | 1.68E-08 | Intergenic variant close to *RAB3C*, *PDE4D* in chromosome 5 | NN and LR |
| **rs6001930** | rs12287832 | 0.68 | 0.09 | 1.11E-09 | Upstream variant proximal to *OVOL1* in chromosome 11 | N/A |
| **rs889312** | rs7842619 | 0.67 | 0.14 | 4.54E-08 | Intron variant in *ANXA13*, *LOC105375739* in chromosome 8 | LR |
| **rs6001930** | rs55910451 | 0.67 | 0.12 | 1.39E-08 | Intron variant in *TASOR2* in chromosome 10 | N/A |
| **rs6001930** | rs1172821 | 0.65 | 0.13 | 3.72E-08 | Intron variant in *BRSK1* in chromosome 19 | LR |
| **rs6001930** | rs7800548 | 0.65 | 0.10 | 4.70E-09 | Intron variant in *FAM185A*, *FBXL13* in chromosome 7 | LR |
| **rs6001930** | rs78540526 | 0.64 | 0.10 | 4.54E-09 | Intergenic variant in chromosome 11 | NN and LR |
| **rs6001930** | rs12406858 | 0.64 | 0.12 | 1.48E-08 | Intron variant in *TENT5C-DT* in chromosome 1 | NN and LR |
| **rs889312** | rs10472076 | 0.63 | 0.14 | 7.84E-08 | Intergenic variant close to *RAB3C*, *PDE4D* in chromosome 5 | NN and LR |
| **rs6001930** | rs149370081 | 0.63 | 0.14 | 8.67E-08 | Intron variant in *RETREG3* in chromosome 17 | LR |
| **rs889312** | rs12406858 | 0.63 | 0.12 | 3.17E-08 | Intron variant in *TENT5C-DT* in chromosome 1 | NN and LR |
| **rs6001930** | rs11242675 | 0.62 | 0.11 | 1.47E-08 | Intergenic variant in chromosome 6 near *FOXQ1* | N/A |
| **rs6001930** | rs4880038 | 0.62 | 0.14 | 8.10E-08 | Intron variant in *PAX5* in chromosome 9 | N/A |
| **rs6001930** | rs17426269 | 0.62 | 0.15 | 1.71E-07 | Intergenic variant in chromosome 1 | N/A |
| **rs889312** | rs7800548 | 0.62 | 0.10 | 5.36E-09 | Intron variant in *FAM185A*, *FBXL13* in chromosome 7 | LR |
| **rs6001930** | rs62334414 | 0.62 | 0.12 | 3.05E-08 | Intron variant in *ADAM29* in chromosome 4 | N/A |
| **rs6001930** | rs187108781 | 0.61 | 0.07 | 1.72E-10 | Intergenic variant in chromosome 5 | N/A |
| **rs889312** | rs1172821 | 0.61 | 0.14 | 1.03E-07 | Intron variant in *BRSK1* in chromosome 19 | LR |
| **rs6001930** | **Menopause** | 0.61 | 0.09 | 1.58E-09 | N/A | N/A |
| **rs6001930** | rs7125780 | 0.60 | 0.13 | 6.64E-08 | Intergenic variant on chromosome 11 | N/A |
| **rs6001930** | rs1428387 | 0.60 | 0.14 | 1.28E-07 | Intron variant in *CEP120* in chromosome 5 | N/A |

Abbreviations: Std = standard error. The interaction between rs6001930 and menopause is in bold. The top two variants, rs6001930 and rs889312, are also in bold. The top interactions all have rs6001930 or rs889312 as "Feature 1", and "Feature 2" is described.

**Table 4: Significant NG-PRS ("gene-by-environment") and NG-NG interactions in the NG+PRS2 neural network model**

| Feature 1 | Feature 2 | Feature 2 Description | Mean Interaction Weight | Std | p-value |
|---|---|---|---|---|---|
| Menopause | rs6001930 | Intron variant in *MRTFA* in chromosome 22 | 0.61 | 0.09 | 1.58E-09 |
| | rs889312 | Intergenic variant near *LOC105378979*, *C5orf67*, *MAP3K1* in chromosome 5 | 0.55 | 0.06 | 1.35E-10 |
| | rs7121616 | Intron variant in *CLMP* in chromosome 11 | 0.38 | 0.09 | 1.52E-07 |
| | Age | N/A | 0.37 | 0.09 | 1.48E-07 |
| Age | rs6001930 | Intron variant in *MRTFA* in chromosome 22 | 0.56 | 0.12 | 5.20E-08 |
| | rs889312 | Intergenic variant near *LOC105378979*, *C5orf67*, *MAP3K1* in chromosome 5 | 0.52 | 0.08 | 5.40E-09 |
| | rs1353747 | Intron variant in *PDE4D* in chromosome 5 | 0.42 | 0.09 | 7.23E-08 |
| | rs12406858 | Intron variant in *TENT5C-DT* in chromosome 1 | 0.41 | 0.09 | 9.59E-08 |
| | rs4880038 | Intron variant in *PAX5* in chromosome 9 | 0.40 | 0.09 | 1.58E-07 |
| | rs11242675 | Intergenic variant in chromosome 6 near *FOXQ1* | 0.39 | 0.08 | 5.11E-08 |
| | rs55910451 | Intron variant in *TASOR2* in chromosome 10 | 0.39 | 0.09 | 1.65E-07 |
| | rs4818836 | Intron variant in *PCNT* in chromosome 21 | 0.39 | 0.09 | 1.49E-07 |
| | rs78540526 | Intergenic variant in chromosome 11 in an enhancer element | 0.40 | 0.09 | 1.39E-07 |
| | rs187108781 | Intergenic variant in chromosome 5 | 0.38 | 0.09 | 1.58E-07 |
| | rs1016578 | Intergenic variant in chromosome 8 | 0.36 | 0.08 | 1.06E-07 |
| Hysterectomy | rs6001930 | Intron variant in *MRTFA* in chromosome 22 | 0.54 | 0.09 | 9.48E-09 |
| Age of Menarche | rs6001930 | Intron variant in *MRTFA* in chromosome 22 | 0.50 | 0.12 | 1.27E-07 |
| | rs1353747 | Intron variant in *PDE4D* in chromosome 5 | 0.42 | 0.08 | 3.15E-08 |
| Alcohol Intake (Frequent) | rs6001930 | Intron variant in *MRTFA* in chromosome 22 | 0.49 | 0.11 | 1.17E-07 |
| | rs1353747 | Intron variant in *PDE4D* in chromosome 5 | 0.40 | 0.09 | 8.33E-08 |

Abbreviations: Std = standard error. Note: The alternative name for *MRTFA* is *MKL1*.

**Table 5: Maximal cliques with NG-PRS interactions in graph of significant pairwise interactions from NG+PRS2 neural network**

| Clique Size | Features |
|---|---|
| 9 | rs6001930, rs1353747, rs889312, rs12406858, rs187108781, rs55910451, rs4818836, **Age**, rs78540526 |
|   | rs6001930, rs1353747, rs889312, rs12406858, rs187108781, rs55910451, rs4818836, **Age**, rs11242675 |
| 5 | rs6001930, rs1353747, rs889312, **Age**, rs4880038 |
| 4 | rs6001930, rs1353747, rs1016578, **Age** |
|   | rs6001930, **Menopause, Age**, rs889312 |
| 3 | rs6001930, rs1353747, **Alcohol Intake (Frequent)** |
|   | rs6001930, rs78540526, **Age** |
|   | rs6001930, rs1353747, **Age of Menarche** |
|   | rs6001930, **Hysterectomy**, rs12287832 |
|   | rs6001930, **Menopause**, rs7121616 |

Abbreviations: NG = non-genetic, PRS = polygenic risk score, PRS-PRS = interaction with only genotype variant features, NG-PRS = interaction with at least one genomic variant and at least one non-genetic feature. Any non-genetic features in these maximal cliques are in bold.

# Supplementary Methods

## Further Details on Cohort Selection

To identify cases and controls, individuals were classified based on whether they had a given ICD-10 billing code in their medical record. Five diagnostic datasets available in the UK Biobank were used to identify cases: primary care records, hospital inpatient records, cancer registry data, death registry data, and self-reported questionnaire data (i.e., patients being asked about their medical history). These were encoded using various systems, including ICD-10, ICD-9, a UK Biobank-designed coding system, and read version 2 and 3 codes. We adapted mappings from each coding system to ICD-10 (**Supplementary Table 13**) and then selected ICD-10 codes that would reasonably identify individuals as being breast cancer cases. Any individual diagnosed with at least one code in the list for breast cancer at any point in time in their medical history would be considered as a prevalent case. To establish cohorts for incident disease risk prediction, we used the year of each participant's first UK Biobank assessment visit as the baseline time point for that participant and excluded all participants with a pre-existing diagnosis prior to their assessment visit. Individuals lost to follow-up in the incident prediction time period, as per data field 191 and the death registry, were excluded from the analyses unless they had an incident diagnosis of the disease.

## Model Training and Selection Details

For each experiment, 10 trials were run systematically, each comprising a random search of hyperparameters. For each trial, random values were selected within specified ranges for each hyperparameter, for a total of 20 possible combinations of hyperparameters, each selection based on a random selection of each hyperparameter. For example, for logistic regression models, the two hyperparameters included learning rate and $L_1$ regularization, and each random selection for each model contained a randomly selected value for each hyperparameter, which was used to train a logistic regression algorithm. After training 20 times as part of the random selection approach, the best-performing model checkpoint was selected based on its validation set performance using area under the curve (AUC). This became the model used for the trial, and this random selection approach was repeated independently for all 10 trials.

## Feature Importance Analysis Using Integrated Gradients – Additional Details

To interpret the neural network models, we calculated input feature importance weights for each model using the integrated gradients framework[1]. After training each model across the ten trials, we ran integrated gradients using the trained model for the test set of each trial to attain feature importance attributions for each participant in the test set (per trial) for each feature. This resulted in ten matrices, each with a feature importance weight for each participant for each feature. For each trial, we then averaged the matrices across the test set to attain an importance value for each feature. This resulted in ten importance weights per feature, one for each trial. The final weights used to interpret each model were the mean (and standard error) of the feature importance weights of the ten trials. To calculate the significance of each feature, we used a one-sample two-tailed t-test comparing the feature importance across the trials to zero and used the Bonferroni correction to adjust for multiple comparisons (alpha = 0.05).

## Feature Category Enrichment for Feature Interactions Analysis

To explore the distribution of interaction weights, we first calculated the mean and standard error of the interaction weight for each feature interaction category (NG-NG, NG-PRS, and PRS-PRS) and then used a one-way analysis of variants (ANOVA) followed by Tukey's honestly significant difference (HSD) to compare these categories.

An additional analysis that we did for the NG+PRS2 neural network model was to explore how enriched each feature category (BC_313, BC_77, NG, or "Both") was for highly weighted (i.e., strong) interactions. Out of the 53.6k unique feature interaction pairs in the NG+PRS2 neural network model, we extracted all pairs in which one or both features in the pair were originally variants in the BC_313 model, and then calculated the mean and standard error of the interaction weight for those feature pairs. We repeated the same for interaction pairs containing at least one BC_77 variant and did the same for NG features and for interactions containing at least one variant out of the 11 that overlapped in both models ("Both" in the table). These groups were not mutually exclusive, given that a feature pair containing one BC_313 variant and on BC_77 variant, for example, would be included in both the BC_313 computations and the BC_77 computations. However, this analysis was designed to provide an estimate of the comparative interaction weights for feature pairs from each respective category. This was done by running a one-way ANOVA using the mean and standard deviation for each category, followed by Tukey's HSD post-hoc statistic.

# Supplementary Figures

**Supplementary Fig. 1: Experimental Design Framework**

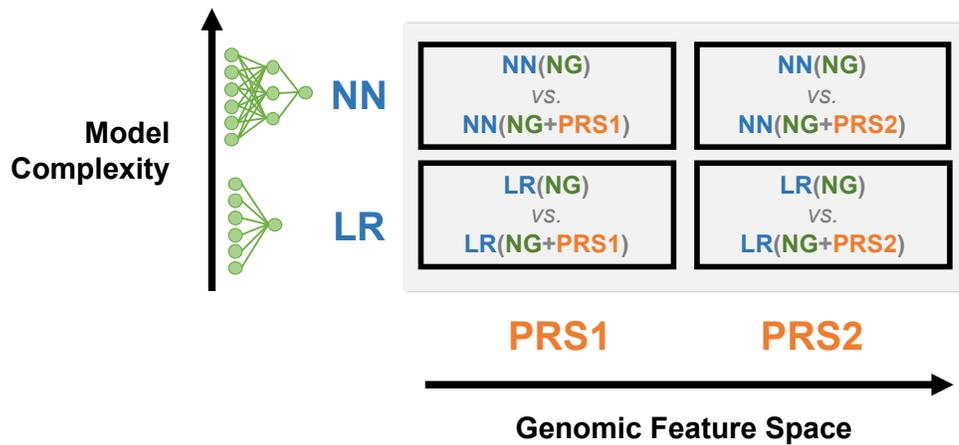

Abbreviations: NN = neural network, LR = logistic regression, PRS1 = feature set with one or more univariate polygenic risk scores, PRS2 = feature set with individual genotype variants. For this study, the "value-add" of each multi-modal model over the non-genetic data only models is calculated as the difference in area under the curve (AUC) between the multi-modal model and the single data modality model for each trial, and then averaged across the trials.

**Supplementary Fig. 2: Precision and recall at various class decision thresholds for single- and multi-modality neural networks and logistic regression models**

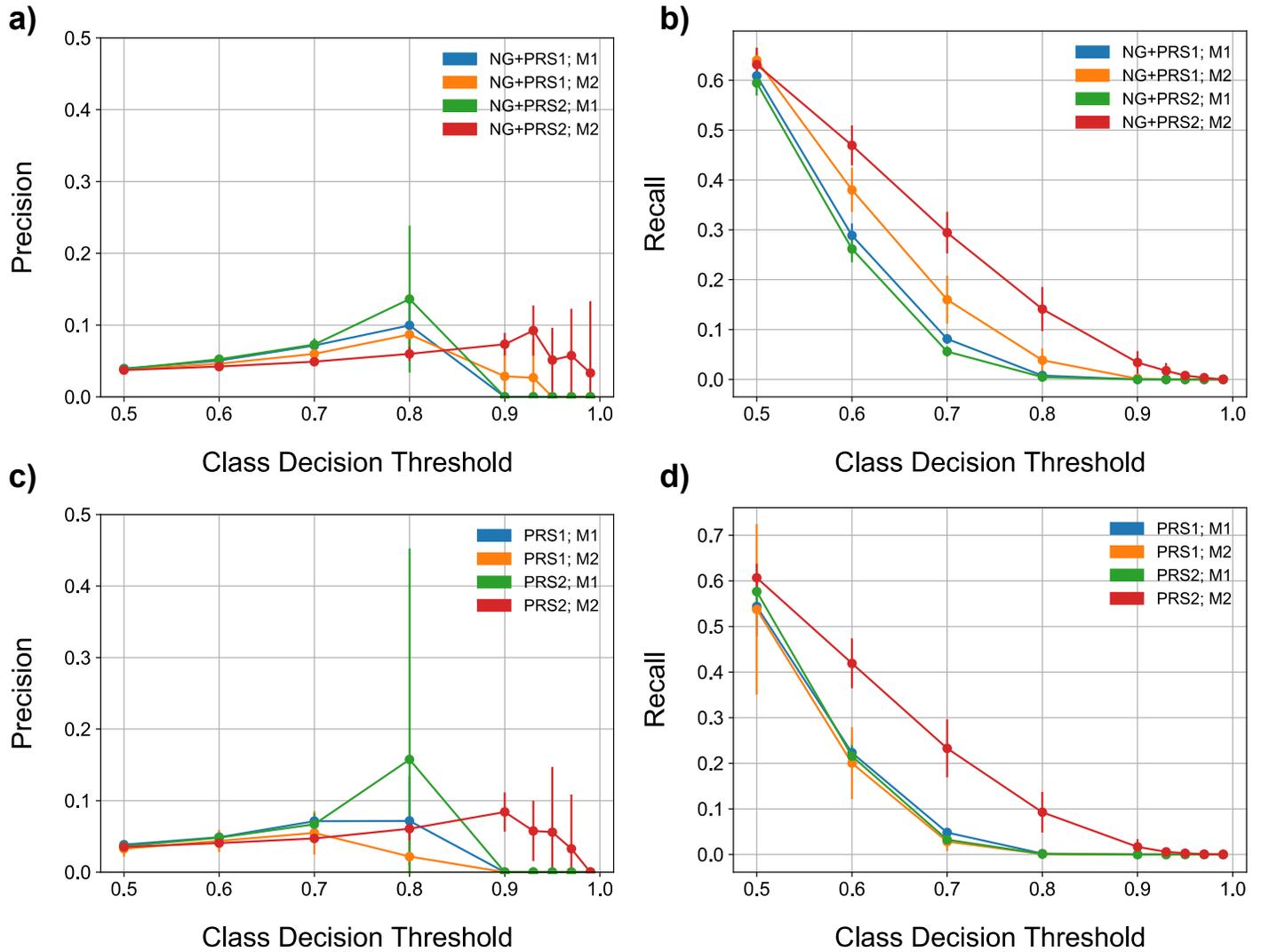

Abbreviations: M1 = model complexity category 1 (logistic regression), M2 = model complexity category 2 (neural network), PRS1 = polygenic risk score category 1 (univariate scores), PRS2 = polygenic risk score category 2 (individual genotype features). **a-b)** Multi-modal NG+PRS1 and NG+PRS2 models. **c-d)** Single-modality polygenic risk models.

**Supplementary Figure 3: NG+PRS2 neural network importance of all significant features with an association with breast cancer score**

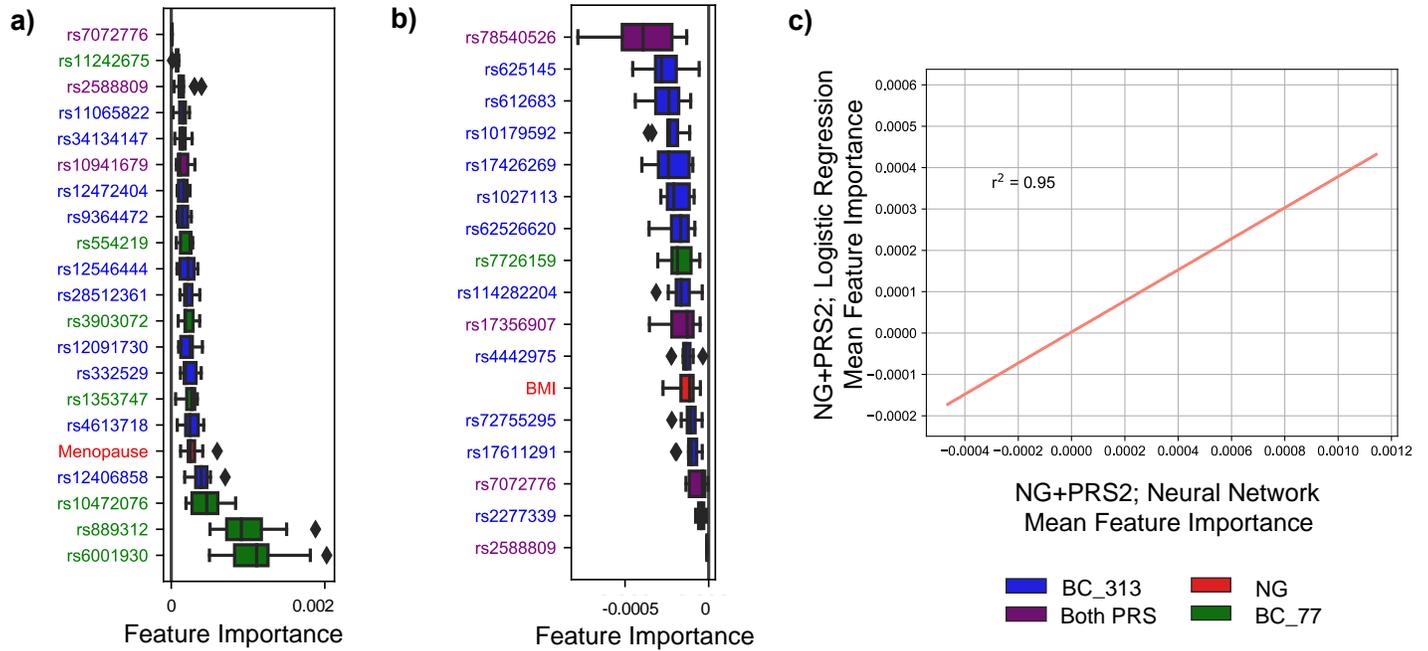

Abbreviations: BMI = body mass index, NG = non-genetic features, PRS2 = feature set with individual genotype variants. Note: There were two control non-genetic risk factor features ("Ever used hormone replacement therapy – prefer not to answer" and "Alcohol intake – no answer") with small importance weights that passed the Bonferroni correction threshold (alpha = 0.05) for identifying features significantly far from zero. These were excluded from the figure above. The figure separates features based on whether they were originally positive or negative in association with the breast cancer risk score. Since the reference allele used for the genotypes is arbitrary, not necessarily the effect allele known to be associated with risk, we focused on the magnitude of importance for the analysis of genotype variants. **a)** Significant features with positive mean importance. The distribution of the feature importance weight (x-axis) across 10 trials is shown, as a boxplot for each feature, with individual points referring to the feature importance value for each trial. **b)** Significant features with negative mean importance. In the NG+PRS2 model, body mass index is negatively associated with risk, and this is significant across the trials. While the magnitude of importance is relatively small, it is significantly greater than zero, suggesting potential model convergence at local minima. **c)** Scatterplot and Pearson correlation of mean feature importance between NG+PRS2 neural network and logistic regression models ($r^2 = 0.95$). The trend of feature importance for the multi-modal models does not change significantly regardless of whether the model is trained using a neural network or logistic regression.

**Supplementary Figure 4: Distribution of feature interaction weights and p-values for multi-modal NG+PRS2 neural network model**

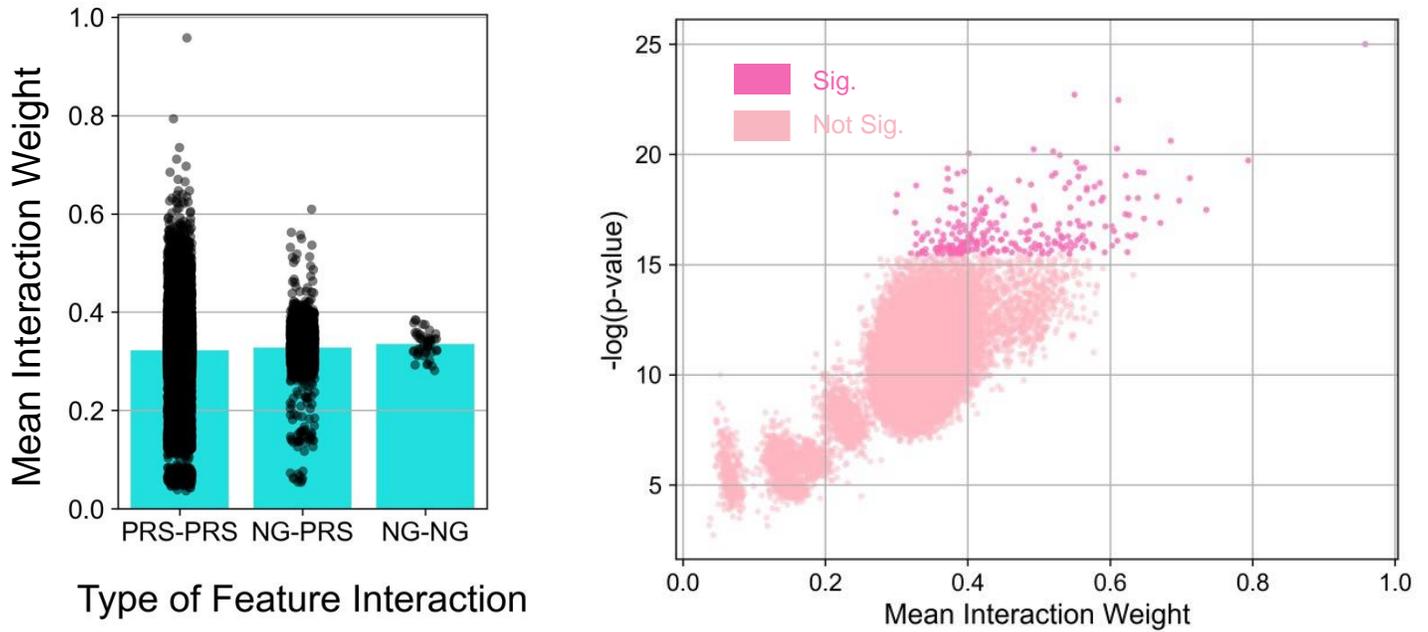

**Left Figure.** Bar plot with points specifying the mean interaction weight of each of ~53.6k possible feature pairs between two non-genetic features (NG-NG), two genotype variants (PRS-PRS), and one non-genetic feature and a genotype variant (NG-PRS). On average, all pairwise interaction weights are larger than zero, but only 239 of these interactions (less than 1%) are significantly larger than zero. **Right Figure.** A scatterplot of the mean pairwise interaction weight (i.e., the same as shown on the y-axis of the left figure) and the negative log of the p-value. In general, feature pairs with a higher mean interaction weight are also more likely to be significant. The interaction between rs889312 and rs6001930 is an outlier at the top right (i.e., lowest p-value and highest interaction weight). Some of the significant interactions have lower weights, suggesting that these were not as important for model training but are still potentially true interactions captured by the model.

**Supplementary Figure 5: Most important significant feature interactions in NG+PRS2 neural network sorted by lowest p-value**

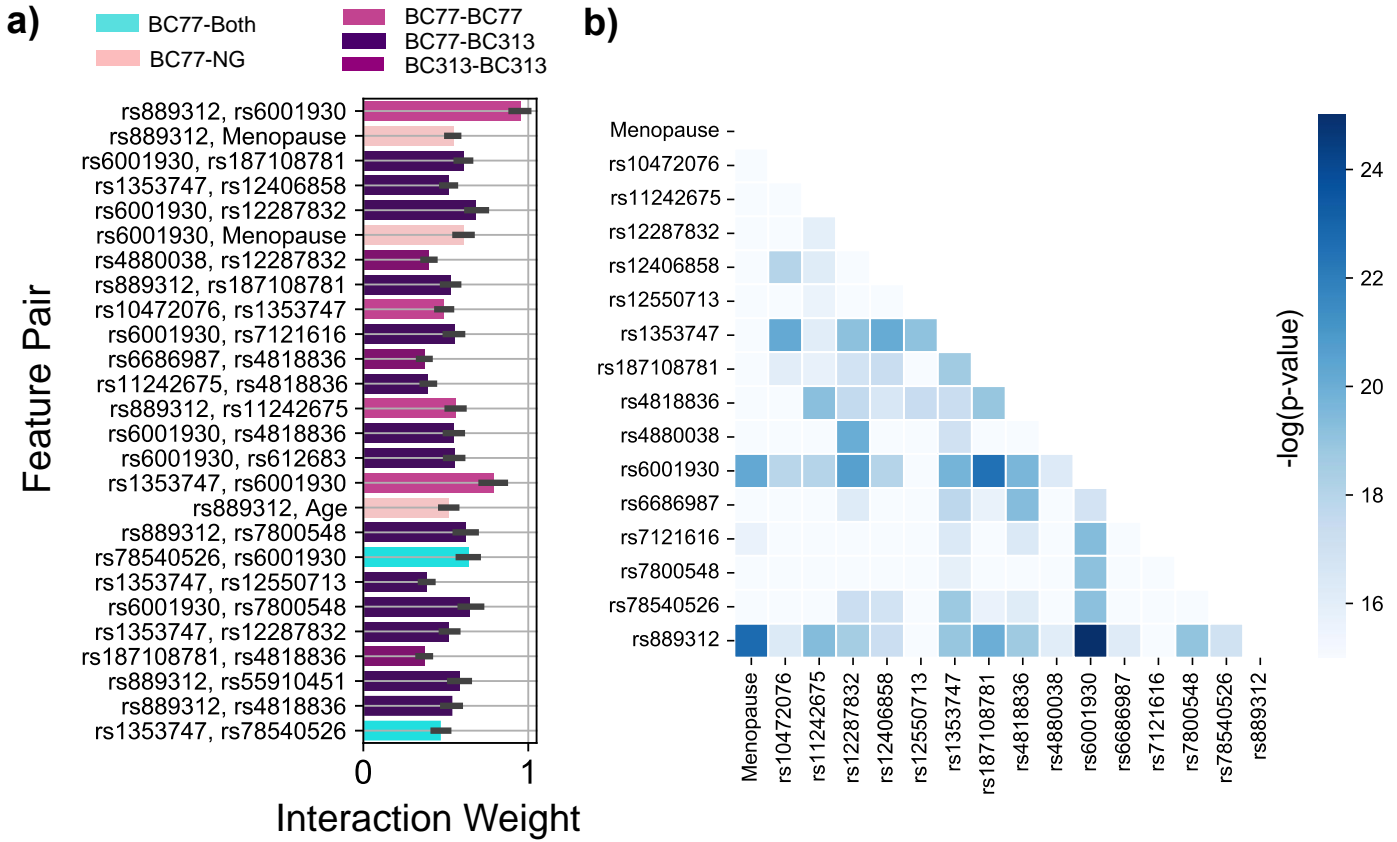

Abbreviations: BC_77 = genotype variants originally selected from the BC_77 polygenic risk score, BC_313 = genotype variants originally selected from the BC_313 polygenic risk score, NG = non-genetic risk factors. **a)** The top 25 most significant pairwise interactions (i.e., with lowest p-values) in the NG+PRS2 multi-modal neural network model, sorted by lowest p-value. The interactions between menopause and rs889312, between menopause and rs6001930, and between age and rs889312, have relatively low p-values, despite the latter two not being in the top-25 when sorted by mean interaction weight. These interaction pairs might represent true interactions captured by the model, but regardless, are less important for disease prediction. **b)** The top 20 most significant pairwise interactions were identified and then the set of unique features in those interactions was compared in a heatmap. As shown, the p-values for the interactions between rs889312 and rs6001930, between menopause and rs889132, and between rs6001930 and rs187108781 (intergenic variant in chromosome 5, ~12 million base pairs away from rs889212), are the most significant. A few others include those between rs6001930 and menopause, between rs6001930 and rs12287832 (upstream variant of *OVOL1* in chromosome 11). Some of these do not include rs6001930 or rs889312, including the interactions between rs12406858 (intron variant in *TENT5C-DT* in chromosome 1) and rs1353747 (intron variant in *PDE4D* in chromosome 5), and between rs12287832 and rs4880038 (intron variant in *PAX5* in chromosome 9).

**Supplementary Figure 6: Heatmaps of top 35 significant features sorted by highest mean interaction weight and lowest p-values, respectively**

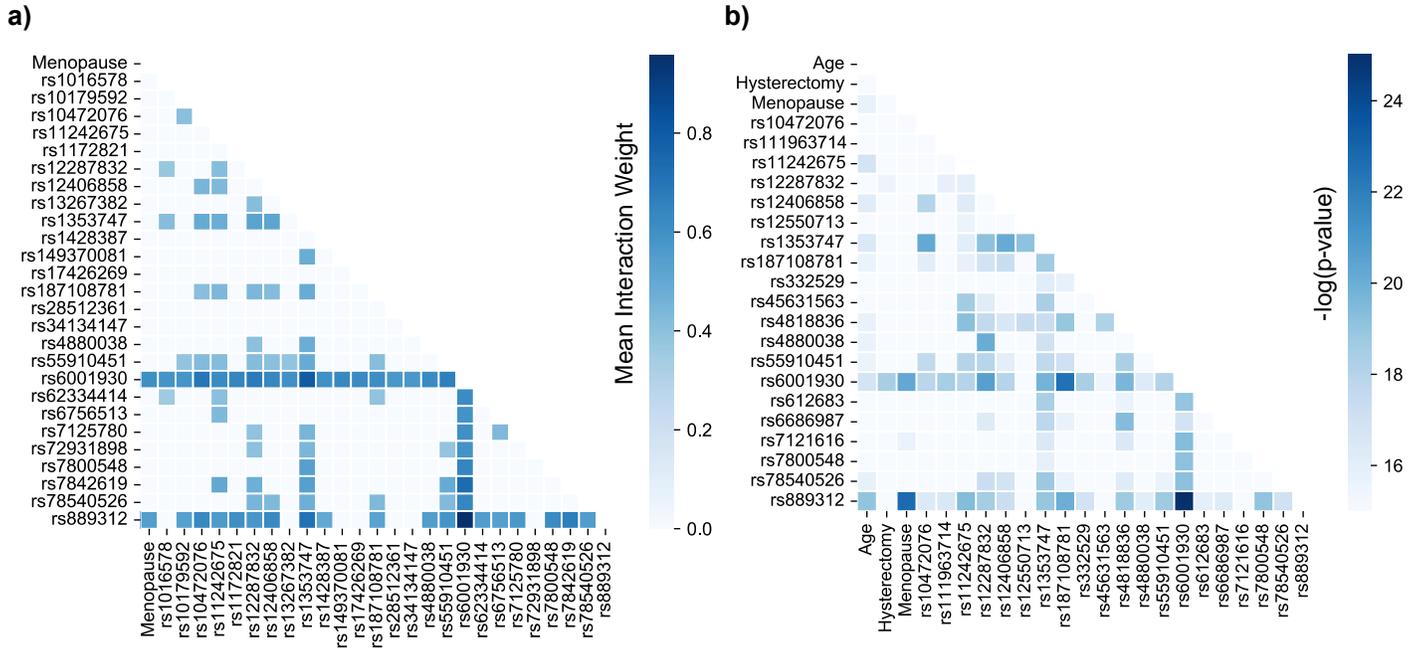

**a)** The top 35 significant interaction pairs with the highest mean weights across the trials are shown, along with all other significant interactions amongst the unique features in these top-35 interaction pairs. All non-significant features are set to zero. The interaction between rs6001930 and rs889312 is the strongest, and all the top 35 strongest interactions contributing towards model performance include one of these. Many features, including menopause, interact with both of those top variants, and some just interact with rs6001930 which is the most important feature in the model. **b)** The top 35 significant interaction pairs with the lowest p-values (i.e., highest negative log p-values) are shown. Again, the interaction between rs6001930 and rs889312 is the most significant. However, others, including non-genetic features such as age, menopause, and whether a participant had a hysterectomy, are some of the most significant interactions. These do not necessarily contribute as much to model performance but may represent actual interactions captured by the model.

**Supplementary Figure 7: Graph of significant features in NG+PRS2 neural network model**

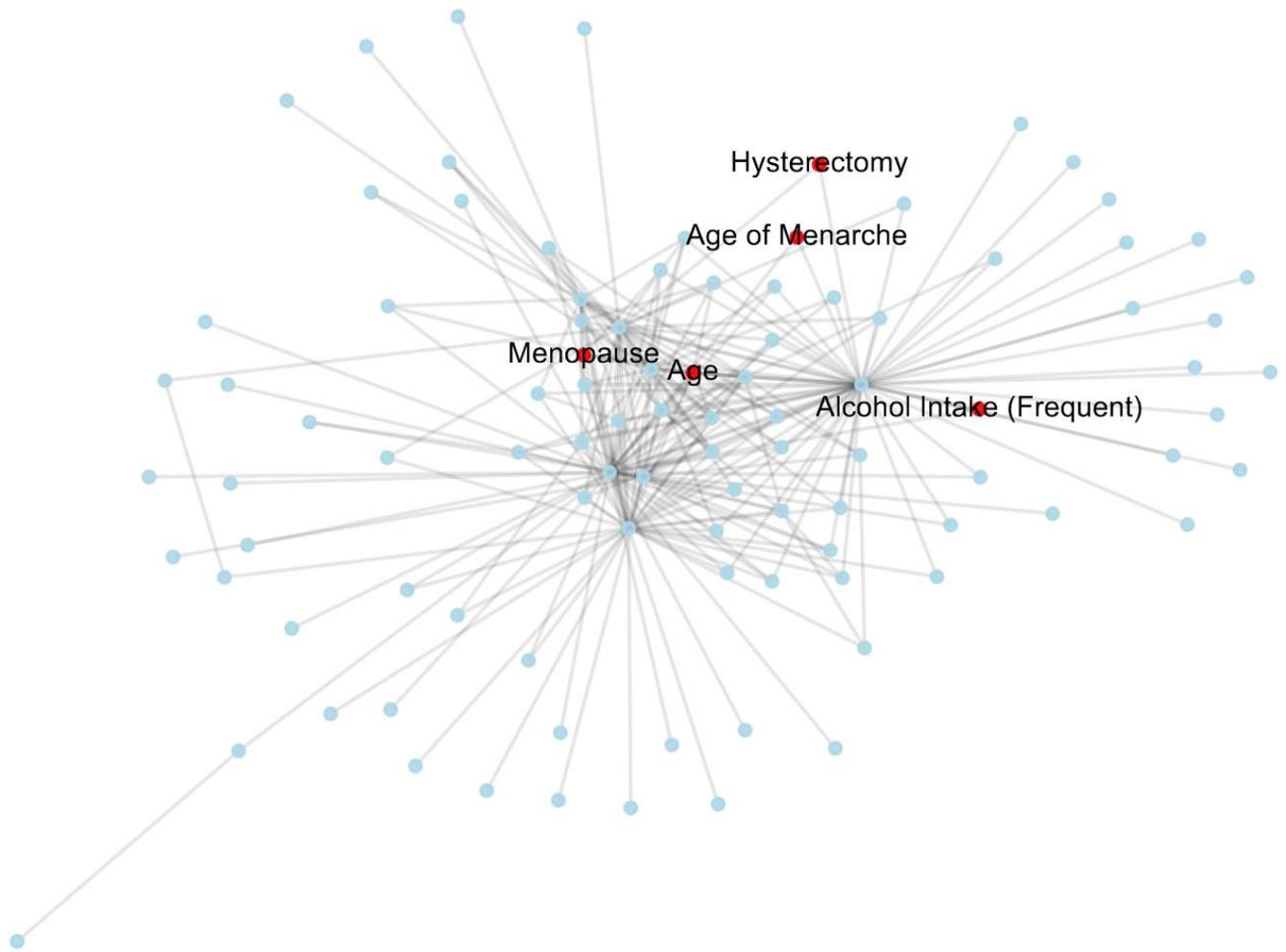

A graph of all 239 significant interactions, and their respective features, is shown, with the non-genetic feature nodes in red, for visualization purposes. Edges are set between any pair of features for the significant interactions. Nodes (i.e., features) that are connected to just one other node, for example, have just one significant pairwise interaction with another feature. Other nodes that have many connections (i.e., rs6001930, which is part of a total of 57 different cliques) are closer to the center.

**Supplementary Figure 8: Graph of top 25 most significant features in NG+PRS2 neural network model, sorted by p-value**

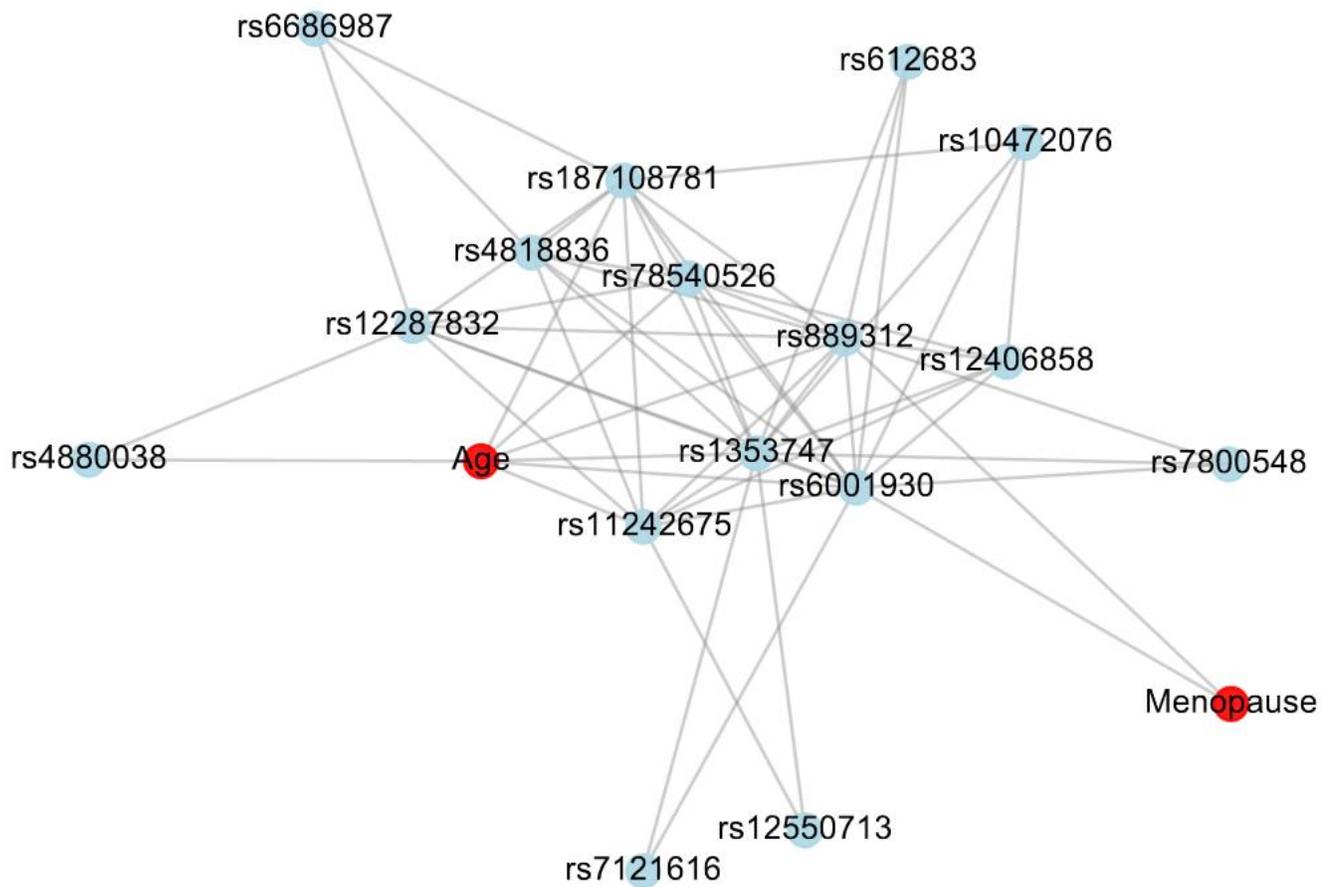

The graph of the top 25 most significant features is shown, for visualization purposes. Age is highly connected to other nodes in the set of most significant features.

**Supplementary Figure 9: Breast Cancer Univariate Polygenic Risk Scores Percentile vs. Prevalence**

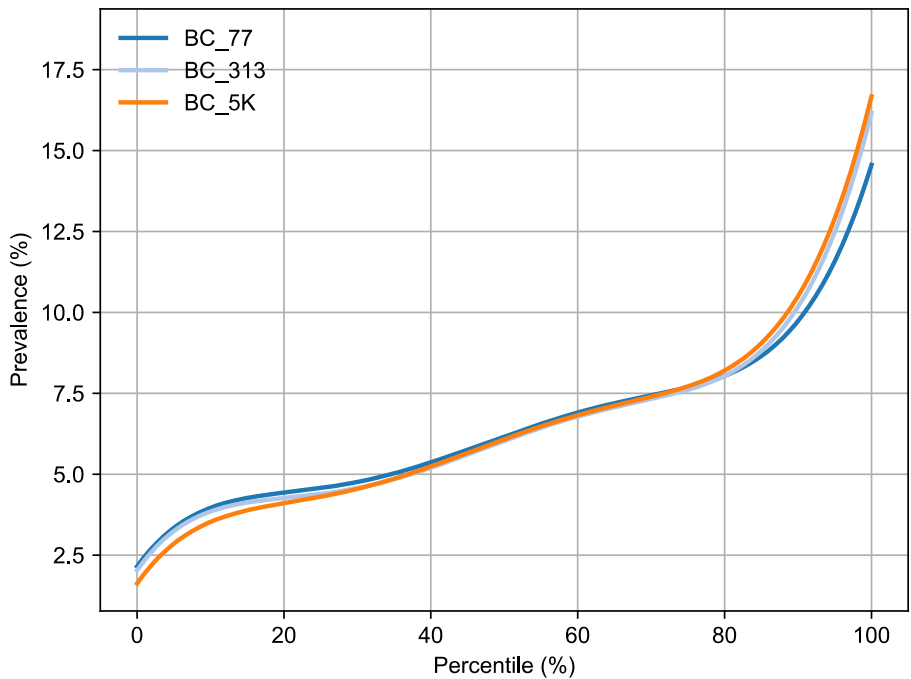

Two of the scores shown (BC_77 and BC_313) were used in the comparative 10-year incident prediction analysis as PRS1. The other score (BC_5k) was calculated as a univariate score and used as an independent comparison for the initial benchmarking of univariate polygenic risk scores. The plot shows the prevalence of breast cancer at each percentile bin of each respective score across 10 trials as individual points (one per trial) and a fitted $5^{th}$ order polynomial curve. As shown, the BC_313 and BC_5k achieve similar risk discrimination at the highest percentiles of risk, while BC_5k has slightly better risk discrimination at the lowest percentiles of risk (i.e., there are fewer true cases with a low score). These results are highly concordant with previous literature showing that individuals in the top 10% of the BC_313 risk score have two-fold higher prevalence than those in the middle 20% of the score[2], [3].

# Supplementary Tables

**Supplementary Table 1: Precision and recall at various class decision thresholds for multi-modal and single-modality models**

| Metric | Threshold | NG+PRS1 | | NG+PRS2 | | PRS1 | PRS2 |
|---|---|---|---|---|---|---|---|
| | | LR | NN | LR | NN | LR | NN |
| Recall (SE) | 0.5 | **0.609 (0.022)** | **0.640 (0.025)** | **0.595 (0.025)** | **0.631 (0.034)** | 0.537 (0.187) | 0.607 (0.031) |
| | 0.6 | **0.289 (0.024)** | **0.380 (0.044)** | **0.262 (0.003)** | **0.469 (0.040)** | **0.200 (0.079)** | **0.419 (0.055)** |
| | 0.7 | **0.081 (0.010)** | **0.160 (0.048)** | **0.056 (0.010)** | **0.294 (0.042)** | **0.029 (0.021)** | **0.233 (0.064)** |
| | 0.8 | **0.008 (0.003)** | **0.038 (0.024)** | **0.005 (0.003)** | **0.141 (0.044)** | **0.001 (0.002)** | **0.093 (0.044)** |
| Precision (SE) | 0.5 | 0.039 (0.002) | 0.038 (0.002) | 0.039 (0.002) | 0.037 (0.002) | 0.033 (0.011) | 0.035 (0.002) |
| | 0.6 | **0.051 (0.004)** | **0.046 (0.004)** | **0.053 (0.005)** | **0.042 (0.002)** | 0.044 (0.015) | 0.041 (0.004) |
| | 0.7 | **0.072 (0.007)** | **0.060 (0.008)** | **0.073 (0.009)** | **0.049 (0.004)** | 0.055 (0.030) | 0.047 (0.005) |
| | 0.8 | 0.100 (0.041) | 0.087 (0.020) | **0.136 (0.102)** | **0.060 (0.010)** | **0.022 (0.035)** | **0.061 (0.016)** |

Abbreviations: AUC = area under the curve, SE = standard error, NG = non-genetic features, PRS1 = polygenic risk scores version 1 (combined univariate scores), PRS2 = polygenic risk scores version 2 (individual variants), LR = logistic regression, NN = neural network. For the columns with multi-modal models, values in bold are those for which the results for the multi-modal NG+PRS1 and NG+PRS2 models, respectively, are significantly different between their logistic regression and neural network counterparts. For the columns with single modality (PRS1 or PRS2) neural network models, values in bold indicate that there is a significant difference between the PRS2 and PRS1 model. This was calculated using an unpaired two-sample t-test at alpha = 0.05 with the mean and standard deviation across the 10 trials. Note: For precision at threshold 0.5 between M1 and M2 for NG+PRS2, the t-test results were calculated using the non-rounded numbers and resulted in a p-value of 0.08. Note: At class decision thresholds over 0.5, NG+PRS2; M2 has significantly higher recall than PRS2; M2. It also has significantly improved precision at class decision threshold 0.5.

**Supplementary Table 2: Comparison of value-add in precision and recall between NG+PRS1 and NG+PRS2 neural network models over the single-modality non-genetic (NG) data neural network models at class decision threshold 0.90**

| Performance Metric | NG+PRS1; NN vs. NG; NN (SE) | NG+PRS2; NN vs. NG; NN (SE) | t statistic | p-value |
|---|---|---|---|---|
| Δ Recall | 0.0015 (0.0027) | 0.0339 (0.0241) | 4.233 | 0.0005 |
| Δ Precision | 0.0286 (0.0547) | 0.0734 (0.0166) | 2.474 | 0.0236 |

Abbreviations: SE = Standard Error. For each pair of models (multi-modality NG+PRS and single-modality NG), the per-trial difference in recall and precision results was calculated at class decision threshold 0.9, and the mean and standard error were then calculated across the 10 trials based on this difference to arrive at each Δ Recall and Δ Precision result. A t-test was then run to compare the Δ Recall and Δ Precision values attained at decision threshold 0.9 for each respective multi-modal model.

**Supplementary Table 3: Count of significant features for each model for 10-year incident breast cancer risk prediction**

| Model Complexity | NG Features | PRS Features | | Multi-Modal Features | |
|---|---|---|---|---|---|
| | | PRS1[1] | PRS2 | NG+PRS1 | NG+PRS2 |
| M1 | 4 | 2 | 12 | 6 | 77 |
| M2 | 4 | 2 | 36 | 6 | 38 |

[1] The PRS1 features are considered significant because they were the only features in the model

**Supplementary Table 4: Overlapping significant features amongst NG+PRS2 and PRS2 logistic regression and neural network models for 10-year incident breast cancer risk prediction**

| Overlapping Group | Feature | Original PRS | Details |
|---|---|---|---|
| NG+PRS2 (LR) NG+PRS2 (NN) PRS2 (LR) PRS2 (NN) | rs10941679 | Both PRS | Intergenic variant close to *MRPS30* |
| | rs17356907 | Both PRS | Intergenic variant close to *USP44* and *PGAM1P5* |
| | rs2588809 | Both PRS | Intron variant in *RAD51B* |
| | **rs78540526** | **Both PRS** | Intergenic variant in chromosome 11 |
| NG+PRS2 (LR) PRS2 (LR) | rs11249433 | Both PRS | Intron variant in *EMBP1* linked to *NOTCH2* gene expression |
| | rs941764 | Both PRS | Intron variant in *CCDC88C*, Downstream of *LOC107984673* |
| NG+PRS2 (NN) PRS2 (NN) | rs332529 | BC_313 | Intergenic variant in chromosome 5 |
| | rs34134147 | BC_313 | Intergenic variant in chromosome 22 |
| | **rs6001930** | **BC_77** | Intron variant in *MRTFA* |
| | rs7072776 | Both PRS | Intron variant in *LOC107984214*, Downstream of *MLLT10* |
| NG+PRS2 (LR) NG+PRS2 (NN) | BMI | N/A | N/A |
| | Experienced Menopause | N/A | N/A |
| | rs10179592 | BC_313 | Intergenic variant in chromosome 2 |
| | **rs10472076** | **BC_77** | Intergenic variant near *RAB3C*, *PDE4D* |
| | rs11065822 | BC_313 | Intron variant in *CUX2* |
| | **rs12406858** | **BC_313** | Intron variant in *TENT5C-DT* |
| | rs12546444 | BC_313 | Intron variant in *ZFPM2* |
| | **rs1353747** | **BC_77** | Intron variant in *PDE4D* |
| | rs17426269 | BC_313 | Intergenic variant in chromosome 1 |
| | rs28512361 | BC_313 | Intron variant in *LOC105373071*, *LOC107985535* |
| | **rs332529** | **BC_313** | Intergenic variant in chromosome 5 |
| | rs4442975 | BC_313 | Intron variant in *IGFBP-AS1* |
| | **rs4613718** | **BC_313** | Intergenic variant in chromosome 5 |
| | rs554219 | BC_77 | Intergenic variant in chromosome 11 (located within ~200 bp from intergenic variant **rs78540526**) |
| | **rs6001930** | **BC_77** | Intron variant in *MRTFA* |
| | **rs612683** | **BC_313** | Intron and noncoding RNA variant in *CDC14A*, *LOC105379827* |
| | **rs625145** | **BC_313** | Intron variant in *SIK3* |
| | rs7726159 | BC_77 | Intron variant in *TERT* |
| | **rs889312** | **BC_77** | Intergenic variant near *LOC105378979*, *C5orf67*, *MAP3K1* (located within ~2 million bp from **rs10472076** and **rs1353747**, variants near *PDE4D*) |
| | rs17356907 | Both PRS | Intergenic variant in chromosome 12 |
| PRS2 (LR) PRS2 (NN) | rs12422552 | Both PRS | Intergenic variant near *GRIN2B*, *ATF7IP* |
| | rs616488 | Both PRS | Intron variant in *PEX14* |
| | rs639355 | BC_313 | Intron variant in *COL8A1* |
| | rs7072776 | Both PRS | Intron variant in *LOC107984214*, Downstream of *MLLT10* |
| | rs9693444 | Both PRS | Intergenic variant near *DUSP4*, *LINC00589* |
| | rs11249433 | Both PRS | Intron variant in *EMBP1* linked to *NOTCH2* gene expression |

Note: Values in bold represent features in the top 10 based on feature importance weight magnitude for the NG+PRS2 neural network model. All of the top 10 neural network NG+PRS2 features are also significant in the NG+PRS2 logistic regression model.

Note: The alternative name for *MRTFA* is *MKL1*.

Note: The top block includes the four variants that are significant for all NG+PRS2 and PRS2 single modality neural network and logistic regression models. These variants are not listed again in the other blocks to avoid redundancy in the table.

**Supplementary Table 5: Top 10 important SNPs in each PRS2 single modality models for 10-year incident breast cancer risk prediction**

| Model | SNP | Position | Original PRS | p-value | Type of Variant | Proximal Gene(s) |
|---|---|---|---|---|---|---|
| NN and LR | **rs10941679** | 5:44706498 | Both PRS | $1.17 \times 10^{-7}$ (NN) $2.62 \times 10^{-6}$ (LR) | Intergenic | *MRPS30* |
| | **rs17356907** | 12:96027759 | Both PRS | $1.69 \times 10^{-7}$ (NN) $2.32 \times 10^{-7}$ (LR) | Intergenic | *USP44, PGAM1P5* |
| | **rs639355** | 3:99403877 | BC_313 | $6.34 \times 10^{-5}$ (NN) $5.07 \times 10^{-5}$ (LR) | Intron | *COL8A1* |
| | **rs12422552** | 12:14413931 | Both PRS | $9.85 \times 10^{-7}$ (NN) $1.41 \times 10^{-7}$ (LR) | Intergenic | *GRIN2B, ATF7IP* |
| NN | **rs2981579** | 10:123337335 | BC_77 | $6.96 \times 10^{-5}$ | Intron | *FGFR2* |
| | **rs910416** | 6:152432902 | BC_313 | $1.11 \times 10^{-5}$ | Intron | *ESR1* |
| | **rs11624333** | 14:68979835 | BC_313 | $7.89 \times 10^{-5}$ | Intron | *RAD51B* |
| | **rs4784227** | 16:52599188 | BC_313 | $4.33 \times 10^{-5}$ | Intron | *CASC16* |
| | **rs6001930** | 22:40876234 | BC_77 | $5.04 \times 10^{-5}$ | Intron | *MRTFA* |
| | **rs851984** | 6:152023191 | BC_313 | $3.63 \times 10^{-5}$ | Intron | *ESR1* |
| LR | **rs11249433** | 1:121280613 | Both PRS | $3.48 \times 10^{-5}$ | Intron | *EMBP1* (affects *NOTCH2* gene expression) |
| | **rs7072776** | 10:22032942 | Both PRS | $2.39 \times 10^{-5}$ | Downstream; Intron | *MLLT10, LOC107984214* |
| | **rs7830152** | 8:143669254 | BC_313 | $1.57 \times 10^{-4}$ | Intergenic | N/A |
| | **rs941764** | 14:91841069 | Both PRS | $1.23 \times 10^{-4}$ | Intron; Downstream | *CCDC88C, LOC107984673* |
| | **rs2588809** | 14:68660428 | Both PRS | $3.08 \times 10^{-5}$ | Intron | *RAD51B* |
| | **rs9693444** | 8:29509616 | Both PRS | $2.58 \times 10^{-6}$ | Intergenic | *DUSP4, LINC00589* |

**Supplementary Table 6: Mean pairwise feature interaction weight for interaction categories in NG+PRS1 and NG+PRS2 neural network**

|         | Feature Interaction Pair Category | Mean (SE)      | Count  |
|---------|-----------------------------------|----------------|--------|
| NG+PRS2 | NG-NG                             | 0.334 (0.026)  | 45     |
|         | NG-PRS                            | 0.327 (0.040)  | 3,180  |
|         | PRS-PRS                           | 0.322 (0.049)  | 50,403 |
| NG+PRS1 | NG-NG                             | 0.124 (0.055)  | 45     |
|         | NG-PRS                            | 0.301 (0.147)  | 20     |
|         | PRS-PRS                           | 0.887 (N/A)    | 1      |

Abbreviations: SE = standard error, NG = non-genetic risk factors, PRS1 = two univariate polygenic risk scores, PRS2 = individual genotype variants.

**Supplementary Table 7: Significant pairwise interactions for NG+PRS1 neural network model**

| Feature 1 | Feature 2 | Mean Interaction Weight | Std | p-value |
|---|---|---|---|---|
| **BC_77** | **BC_313** | **0.89** | **0.19** | **5.28E-08** |
| **BC_313** | **Age** | **0.74** | **0.29** | **1.05E-05** |
| **BC_77** | **Age** | **0.55** | **0.18** | **2.03E-06** |
| **BC_313** | **Family Hx** | **0.48** | **0.18** | **7.59E-06** |
| **BC_313** | **BMI** | **0.40** | **0.21** | **1.01E-04** |
| **BC_77** | **Family Hx** | **0.39** | **0.11** | **5.79E-07** |
| **BC_77** | **BMI** | **0.35** | **0.08** | **1.66E-07** |
| **Family Hx** | **Age** | **0.32** | **0.10** | **1.96E-06** |
| Age | BMI | 0.29 | 0.12 | 1.62E-05 |
| **BC_313** | **Alcohol Intake (Frequent)** | **0.25** | **0.12** | **3.89E-05** |
| **BC_313** | **Alcohol Intake (Never)** | **0.23** | **0.12** | **9.14E-05** |
| **BC_313** | **Age of Menarche** | **0.22** | **0.11** | **8.28E-05** |
| **Family Hx** | **BMI** | **0.22** | **0.06** | **4.46E-07** |
| Age | Alcohol Intake (Frequent) | 0.20 | 0.11 | 1.06E-04 |
| **BC_313** | **HRT** | **0.20** | **0.06** | **7.05E-07** |
| **BC_77** | **HRT** | **0.20** | **0.10** | **8.12E-05** |
| Hysterectomy | Age | 0.17 | 0.09 | 8.24E-05 |
| Age | Age of Menarche | 0.16 | 0.09 | 1.43E-04 |
| Age | Menopause | 0.15 | 0.06 | 1.12E-05 |
| **Family Hx** | **Alcohol Intake (Frequent)** | **0.15** | **0.06** | **1.31E-05** |
| Age | Oral Contraceptive | 0.14 | 0.07 | 6.43E-05 |
| HRT | Age | 0.13 | 0.05 | 9.00E-06 |
| Family Hx | Oral Contraceptive | 0.13 | 0.06 | 2.38E-05 |
| Age of Menarche | BMI | 0.13 | 0.04 | 2.73E-06 |
| BMI | Alcohol Intake (Frequent) | 0.13 | 0.06 | 3.86E-05 |
| Hysterectomy | BMI | 0.12 | 0.05 | 2.16E-05 |
| **Family Hx** | **Age of Menarche** | **0.12** | **0.05** | **6.14E-06** |
| **Family Hx** | **HRT** | **0.11** | **0.06** | **9.33E-05** |
| **Family Hx** | **Menopause** | **0.09** | **0.02** | **2.49E-07** |
| HRT | Alcohol Intake (Frequent) | 0.08 | 0.03 | 1.47E-05 |
| HRT | Alcohol Intake (Never) | 0.07 | 0.03 | 3.98E-05 |

Abbreviations: BMI = body mass index, Hx = history, HRT = hormone replacement therapy. The rows are sorted from highest to lowest mean pairwise interaction weight. Feature pairs with either BC_77, BC_313, or family history are highlighted.

**Supplementary Table 8: Pairwise feature interaction enrichment by feature category for NG+PRS2 neural network**

| Feature Category 1 | Feature Category 2 | Feature 1 Mean (SE), Count | Feature 2 Mean (SE), Count | Difference | Significance |
|---|---|---|---|---|---|
| **BC_313** | **BC_77** | 0.3234 (0.0451) 50,388 | 0.3141 (0.0735) 17,850 | 0.0093 | **Significant. BC_313 > BC_77** |
| **BC_313** | **NG** | 0.3234 (0.0451) 50,388 | 0.3275 (0.0398) 3,225 | -0.0041 | **Significant. NG > BC_313** |
| **BC_313** | **Both** | 0.3234 (0.0451) 50,388 | 0.3147 (0.0459) 3,542 | 0.0087 | **Significant. BC_313 > Both** |
| **BC_77** | **NG** | 0.3141 (0.0735) 17,850 | 0.3275 (0.0398) 3,225 | -0.0134 | **Significant. NG > BC_77** |
| BC_77 | Both | 0.3141 (0.0735) 17,850 | 0.3147 (0.0459) 3,542 | -0.0006 | Not Significant. |
| **NG** | **Both** | 0.3275 (0.0398) 3,225 | 0.3147 (0.0459) 3,542 | 0.0128 | **Significant. NG > Both.** |

Abbreviations: CI = confidence interval, SE = standard error, NG = non-genetic. Note: The difference is calculated between the Feature Category 1 mean interaction weight (across all interaction pairs that include at least one Feature Category 1 feature) and that of Feature Category 2. The "Significance" column indicates whether one category of features is more heavily enriched for higher-weighted interactions with other features in the model, compared to its comparison feature category. NG features tend to be in higher weighted pairwise interactions with other features, on average, compared to all other feature categories. The BC_313 features are similarly enriched for higher-weighted interactions on average. The BC_77 features and those features from both BC_77 and BC_313 scores (i.e., the 11 overlapping variants) are significantly less enriched for highly weighted pairwise interactions on average. However, the BC_77 variants, rs889312 and rs6001930, are outliers and the significant interactions are highly enriched for pairs with at least one of these features.

**Supplementary Table 9: Top 30 significant feature interactions in NG+PRS2 neural network model with the largest mean weights**

| Feature 1 | Feature 2 | Mean Interaction Weight | Std | p-value |
|---|---|---|---|---|
| **rs6001930** | **rs889312** | 0.96 | 0.08 | 1.37E-11 |
| **rs6001930** | rs1353747 | 0.79 | 0.12 | 2.67E-09 |
| **rs6001930** | rs7842619 | 0.73 | 0.14 | 2.52E-08 |
| **rs889312** | rs1353747 | 0.71 | 0.12 | 5.98E-09 |
| **rs6001930** | rs10472076 | 0.70 | 0.13 | 1.68E-08 |
| **rs6001930** | rs12287832 | 0.68 | 0.09 | 1.11E-09 |
| **rs889312** | rs7842619 | 0.67 | 0.14 | 4.54E-08 |
| **rs6001930** | rs55910451 | 0.67 | 0.12 | 1.39E-08 |
| **rs6001930** | rs1172821 | 0.65 | 0.13 | 3.72E-08 |
| **rs6001930** | rs7800548 | 0.65 | 0.10 | 4.70E-09 |
| **rs6001930** | rs78540526 | 0.64 | 0.10 | 4.54E-09 |
| **rs6001930** | rs12406858 | 0.64 | 0.12 | 1.48E-08 |
| **rs889312** | rs10472076 | 0.63 | 0.14 | 7.84E-08 |
| **rs6001930** | rs149370081 | 0.63 | 0.14 | 8.67E-08 |
| **rs889312** | rs12406858 | 0.63 | 0.12 | 3.17E-08 |
| **rs6001930** | rs11242675 | 0.62 | 0.11 | 1.47E-08 |
| **rs6001930** | rs4880038 | 0.62 | 0.14 | 8.10E-08 |
| **rs6001930** | rs17426269 | 0.62 | 0.15 | 1.71E-07 |
| **rs889312** | rs7800548 | 0.62 | 0.10 | 5.36E-09 |
| **rs6001930** | rs62334414 | 0.62 | 0.12 | 3.05E-08 |
| **rs6001930** | rs187108781 | 0.61 | 0.07 | 1.72E-10 |
| **rs889312** | rs1172821 | 0.61 | 0.14 | 1.03E-07 |
| **rs6001930** | **Menopause** | 0.61 | 0.09 | 1.58E-09 |
| **rs6001930** | rs7125780 | 0.60 | 0.13 | 6.64E-08 |
| **rs6001930** | rs1428387 | 0.60 | 0.14 | 1.28E-07 |
| **rs6001930** | rs13267382 | 0.60 | 0.13 | 5.30E-08 |
| **rs6001930** | rs72931898 | 0.59 | 0.12 | 5.37E-08 |
| **rs6001930** | rs6756513 | 0.59 | 0.14 | 1.73E-07 |
| **rs6001930** | rs1016578 | 0.59 | 0.11 | 1.44E-08 |
| **rs6001930** | rs10179592 | 0.59 | 0.11 | 1.67E-08 |

Abbreviations: Std = standard error. The interaction between rs6001930 and menopause is in bold. The top two variants, rs6001930 and rs889312, are also in bold.

**Supplementary Table 10: Top 24 significant feature interactions in NG+PRS2 neural network model with the smallest p-values**

| Feature 1 | Feature 2 | Mean Interaction Weight | Std | p-value |
|---|---|---|---|---|
| **rs6001930** | **rs889312** | 0.96 | 0.08 | 1.37E-11 |
| **rs889312** | **Menopause** | **0.55** | **0.06** | **1.35E-10** |
| **rs6001930** | rs187108781 | 0.61 | 0.07 | 1.72E-10 |
| **rs6001930** | rs12287832 | 0.68 | 0.09 | 1.11E-09 |
| **rs6001930** | **Menopause** | **0.61** | **0.09** | **1.58E-09** |
| rs10472076 | rs1353747 | 0.49 | 0.07 | 1.63E-09 |
| rs12406858 | rs1353747 | 0.52 | 0.07 | 1.76E-09 |
| rs4880038 | rs12287832 | 0.40 | 0.06 | 1.94E-09 |
| rs187108781 | **rs889312** | 0.53 | 0.08 | 2.11E-09 |
| **rs6001930** | rs1353747 | 0.79 | 0.12 | 2.67E-09 |
| **rs6001930** | rs4818836 | 0.55 | 0.08 | 2.94E-09 |
| **rs889312** | rs11242675 | 0.56 | 0.09 | 3.78E-09 |
| rs7121616 | **rs6001930** | 0.56 | 0.09 | 3.82E-09 |
| rs6686987 | rs4818836 | 0.37 | 0.06 | 3.90E-09 |
| rs11242675 | rs4818836 | 0.39 | 0.06 | 4.45E-09 |
| **rs6001930** | rs78540526 | 0.64 | 0.10 | 4.54E-09 |
| rs7800548 | **rs6001930** | 0.65 | 0.10 | 4.70E-09 |
| rs1353747 | rs12287832 | 0.52 | 0.08 | 4.79E-09 |
| rs12550713 | rs1353747 | 0.39 | 0.06 | 4.88E-09 |
| rs7800548 | **rs889312** | 0.62 | 0.10 | 5.36E-09 |
| **Age** | **rs889312** | **0.52** | **0.08** | **5.40E-09** |
| **rs6001930** | rs612683 | 0.55 | 0.09 | 5.53E-09 |
| **rs889312** | rs1353747 | 0.71 | 0.12 | 5.98E-09 |
| rs187108781 | rs4818836 | 0.37 | 0.06 | 6.09E-09 |

Abbreviations: Std = standard error. The interaction between rs6001930 and menopause is in bold. The top two variants, rs6001930 and rs889312, are also in bold.

**Supplementary Table 11: Maximal cliques of at least five features in graph of significant pairwise interactions from NG+PRS2 neural network**

| Clique Size | Features | Interaction Type |
|---|---|---|
| 9 | **rs6001930, rs1353747, rs889312**, rs12406858, rs187108781, **rs55910451**, rs4818836, **Age**, **rs78540526** | **NG-PRS** |
|   | **rs6001930, rs1353747, rs88931**2, rs12406858, rs187108781, **rs55910451**, rs4818836, **Age**, rs11242675 |  |
| 8 | **rs6001930, rs1353747, rs889312**, rs12287832, rs4818836, rs187108781, **rs55910451, rs78540526** | PRS-PRS |
|   | **rs6001930, rs1353747, rs889312**, rs12287832, rs4818836, rs187108781, **rs55910451**, rs11242675 |  |
|   | **rs6001930, rs1353747, rs889312**, rs12287832, rs4818836, rs7842619, rs11242675, **rs55910451** |  |
| 7 | **rs6001930, rs1353747, rs889312**, rs12287832, rs4818836, rs187108781, rs6686987 | PRS-PRS |
|   | **rs6001930, rs1353747, rs889312**, rs12406858, rs187108781, **rs55910451**, rs10472076 |  |
| 6 | **rs6001930**, rs62334414, rs187108781, **rs889312**, rs11242675, rs4818836 | PRS-PRS |
|   | **rs6001930, rs1353747, rs889312**, rs12287832, rs4818836, rs1027113 |  |
|   | **rs6001930, rs1353747**, rs45631563, rs4818836, rs11242675, rs12287832 |  |
| 5 | **rs6001930**, rs625145, **rs889312**, rs11242675, rs4818836 | PRS-PRS |
|   | **rs6001930, rs1353747**, rs72931898, **rs55910451**, rs12287832 |  |
|   | **rs6001930, rs1353747, rs889312**, rs12287832, rs7125780 |  |
|   | **rs6001930, rs1353747, rs889312**, rs12287832, rs4880038 |  |
|   | **rs6001930, rs1353747, rs889312**, rs12287832, rs11205303 |  |
|   | **rs6001930, rs1353747, rs889312**, rs6030585, **rs55910451** |  |
|   | **rs6001930, rs1353747, rs889312**, rs332529, rs187108781 |  |
|   | **rs6001930, rs1353747, rs889312**, rs612683, rs4818836 |  |
|   | **rs6001930, rs1353747, rs889312, Age**, rs4880038 | **NG-PRS** |
|   | **rs6001930, rs1353747, rs889312,** rs111458676, **rs55910451** | PRS-PRS |
|   | **rs6001930**, rs10179592, **rs889312**, rs10472076, **rs55910451** |  |

Abbreviations: NG = non-genetic, PRS = polygenic risk score, PRS-PRS = interaction with only genotype variant features, NG-PRS = interaction with at least one genomic variant and at least one non-genetic feature. The variants, rs6001940, rs889312, rs1353747, rs55910451, and rs78540526, and any non-genetic features in these maximal cliques, are in bold.

**Supplementary Table 12:** Number of maximal clique memberships per feature in graph of significant pairwise interactions from NG+PRS2 neural network

| Feature | Number of Clique Memberships | Description |
| --- | --- | --- |
| rs6001930 | 57 | Intron variant in *MRTFA* in chromosome 5 |
| rs1353747 | 47 | Intron variant in *PDE4D* in chromosome 5 |
| rs889312 | 37 | Intergenic variant near *LOC105378979*, *C5orf67*, *MAP3K1* in chromosome 5 |
| rs12287832 | 21 | Upstream variant proximal to *OVOL1* in chromosome 11 |
| rs481836 | 17 | Intergenic variant in chromosome 10 |
| rs11242675 | 16 | Intergenic variant in chromosome 6 near *FOXQ1* |
| rs55910451 | 15 | Intron variant in *TASOR2* in chromosome 10 |
| rs187108781 | 8 | Intergenic variant in chromosome 5 |
| rs12406858 | 6 | Intron variant in *TENT5C-DT* in chromosome 1 |
| Age | 5 | N/A |

Note: Only the top 10 features with the greatest number of clique memberships are shown. Note: The alternative name for *MRTFA* is *MKL1*.

**Supplementary Table 13: Code mappings to ICD-10**

| Clinical Code Mappings | Original Mapping System(s) |
|---|---|
| UKB Data Coding 6 to ICD-10 | UKB provides a one-to-one mapping ("data coding 609") |
| Read Version 2 (V2) and Clinical Terms Version 3 (CTV3) to ICD-10 | UKB data coding 1834 mapping from Read V2 to three-character ICD-10 codes |
| | UKB data coding 1835 mapping from Read CTV3 to three-character ICD-10 codes |
| | The 2018 release of Read CTV3 to ICD-10 mapping provided by TRUD website from NHS Digital |
| ICD-9 to ICD-10 | UKB data coding 1836 mapping from ICD-9 to ICD-10 |
| | The 2018 US Centers for Medicare and Medicaid Services (CMS) General Equivalence Mapping (GEM) from ICD-9-CM to ICD-10-CM |

Abbreviations: UKB = UK Biobank, NHS = National Health Service, ICD = International Classification of Disease, CM = Clinical Modification, CTV3 = Clinical Terms Version 3, TRUD = Technology Reference Update Distribution

**Supplementary Table 14: ICD-10 codes used to define incident breast cancer cases and controls for breast cancer**

| ICD-10 Codes | Description |
|---|---|
| C50 (C50.0, C50.1, C50.2, C50.3, C50.4, C50.5, C50.6, C50.8, C50.9) | Malignant neoplasm of breast, including nipple and areola, central portion of breast, upper-inner, lower-inner, upper-outer, or lower-outer quadrant of breast, axillary tail of breast, etc. |

**Supplementary Table 15: Case and control counts for breast cancer and psoriasis for 10-year incidence prediction**

| Disease | 10-Year Incident Cases[1] | Controls | Percent Cases (%) |
|---|---|---|---|
| **Breast Cancer** | 5,174 | 188,043 | 2.68% |

[1] The 10-year incident case counts are lower than the overall prevalence for each disease in the UK Biobank population because we removed individuals lost to follow-up in the 10-year window, individuals who opted out of the study, individuals with a pre-existing diagnosis of the disease before the time of their first assessment visit, any individuals that were not of White British ancestry, anyone without any available genomic data or diagnostic data prior to their first assessment visit, and anyone diagnosed with the disease after the 10-year window. For example, while as many as 17,600 individuals out of in the UK Biobank population were classified as having a prevalent diagnosis of breast cancer, only 5,174 (~29%) of the modified UK Biobank sample subset were considered as incident cases. Likewise, as many as 13,120 individuals in the UK Biobank dataset had a diagnosis of psoriasis, but only 1,184 (~9%) were 10-year incident cases. Because of the filtering based on sex and pre-existing diagnoses, each disease had a different number of total samples.

[2] For Analysis 2, the baseline univariate polygenic risk score and the gene embeddings were generated using the train and validation sets after splitting the overall dataset into train, validation, and test in a 70/20/10 ratio. For this reason, only the test set was used for the 10-year incident disease prediction tasks.

**Supplementary Table 16: Breast Cancer (BC) clinical features for non-genetic feature set 1 (NG1)**

| Risk Factor Category | Feature | UK Biobank Data Field | Data Type | Number of Features | Description |
|---|---|---|---|---|---|
| Baseline Demographics | Age at Recruitment | 21022 | Numerical | 1 | N/A |
| Physical Measurements | Body Mass Index (BMI) | 21001 | Numerical | 1 | Measured in kg/m$^2$ |
| Female-Specific Factors | Age of Menarche | 2714 | Numerical and Binary | 3 | Age of menarche is a numerical feature and null values were imputed. "Do not know" and "Prefer not to answer" were options in the questionnaire and included as binary control features. |
| | Experienced Menopause | 2724 | Binary | 5 | Questionnaire options included "Yes", "No", "Do not know", "Do not know – had hysterectomy", and "Prefer not to answer" |
| | Ever used hormone replacement therapy (HRT) | 2814 | Binary | 4 | Questionnaire options included "Yes", "No", "Prefer not to answer", and "Do not know" |
| | Ever used oral contraceptive | 2784 | Binary | 4 | Options included "Yes", "No", "Prefer not to answer", and "Do not know" |
| Lifestyle Factors | Alcohol Intake Frequency | 1558 | Binary | 7 | Collected using touchscreen questionnaire. The options included: "Daily or Almost Daily", "Never", "1-2x per week", "1-3x per month", "3-4x per week", "Special Occasions Only", and "Prefer not to answer" |
| Family History | Mother Diagnosed with Breast Cancer | 20110 | Binary | 1 | Collected using a touchscreen questionnaire |
| | Sibling Diagnosed with Breast Cancer | 20111 | Binary | 1 | Collected using a touchscreen questionnaire |

Abbreviations: NG = non-genetic. All data points were collected at the time of each individual's recruitment to the UK Biobank. Missing numerical features were imputed using the mean of all non-missing values for each feature. The features were scaled for training. To avoid excluding individuals with "Prefer Not to Answer" for the alcohol intake frequency question, this was included as a binary feature. It was also included to serve as a control feature to validate findings, as it does not contain relevant data for risk prediction. The same was done for the female-specific factors, including age of menarche, history of use of hormone replacement therapy or oral contraceptives, and history of menopause.

**Supplementary Table 17: Feature counts for each feature set**

| Data Modality | Feature Set | Number of Features |
|---|---|---|
| Genomic | PRS1 | 2 |
|  | PRS2 | 318 |
| Non-Genetic | NG | 27 |
| Multi-Modal | NG+PRS1 | 29 |
|  | NG+PRS2 | 345 |

**Supplementary Table 18: Breast Cancer (BC) polygenic risk scores selected from literature**

| PRS | AUC (SE)[1] | Description |
|---|---|---|
| **BC_77 [4]** | 0.578 (0.005) | A total of 77 variants identified from the BCAC Collaborative Oncological Gene-Environmental Study (COGS), mostly with p-values $< 5 \times 10^{-8}$ for association with breast cancer |
| **BC_313 [3]** | 0.586 (0.007) | A total of 313 variants selected using p-value thresholding combined with forward stepwise regression across 1 million base pair regions of the genome, followed by joint logistic regression on the selected variants to get weights (i.e., log odds ratios). The analyses were done using data from 94,075 breast cancer cases and 75,017 controls of European ancestry from 69 studies as part of BCAC. |
| **BC_5k[2] [5]** | 0.591 (0.006) | A total of 5,218 variants and their weights, selected using LDPred. [6] |

Abbreviations: LD = linkage disequilibrium, GWAS = genome-wide association study, BCAC = Breast Cancer Association Consortium, AUC = area under the precision recall curve, SE = standard error

[1] The AUC is calculated on all samples with available genomic data for prevalent breast cancer prediction.

[2] BC_5k was used for initial benchmarking of the univariate polygenic risk scores but was not included in the 10-year incident prediction analysis.

**Supplementary Table 19: Non-genetic features adapted for the interactions interpretation analysis**

| Feature Concept | Features | Adaptation for Interactions Analysis |
|---|---|---|
| Age | Age | **Age** |
| BMI | BMI | **BMI** |
| Age of Menarche | Age of Menarche | **Age of Menarche** |
| | Prefer not to answer | Removed |
| Experienced Menopause | Yes | **Menopause** |
| | No | Removed |
| | Do not know | |
| | Prefer not to answer | |
| | Not sure – had hysterectomy | **Hysterectomy** |
| Ever used HRT | Yes | **HRT** |
| | No | Removed |
| | Prefer not to answer | |
| | Do not know | |
| Ever used oral contraceptive | Yes | **Oral Contraceptive** |
| | No | Removed |
| | Prefer not to answer | |
| | Do not know | |
| Alcohol Intake Frequency | Never | **Alcohol Intake (Never)** |
| | Daily or almost daily | **Alcohol Intake (Frequent)** |
| | 3-4x per week | |
| | 1-3x per month | Removed |
| | 1-2x per week | |
| | 3-4x per week | |
| | Special Occasions Only | |
| | Prefer not to answer | |
| Family history | Mother | **Family History** |
| | Sibling | |

Abbreviations: BMI = body mass index, HRT = hormone replacement therapy

Note: For alcohol intake frequency, "special occasions only", "1-2x per week", and "3-4x per month" were removed to keep the answers at the extreme (Never, and Frequent), to simplify the model interpretation.